\documentclass[prb,twocolumn,showpacs,superscriptaddress]{revtex4-1}
\usepackage{graphicx}
\usepackage{amsmath,amssymb,amstext,amsthm}
\usepackage{bbold}
\usepackage{cancel}
\usepackage{color}
\usepackage{verbatim}
\usepackage{booktabs}
\usepackage[caption=false]{subfig}

\begin{document}

\title{Topological surface states and Andreev bound states in
superconducting iron pnictides}

\author{Alexander Lau}
\affiliation{Institute for Theoretical Solid State Physics, IFW Dresden,
01171 Dresden, Germany}
\affiliation{Institute of Theoretical Physics, Technische Universit\"at
Dresden, 01062 Dresden, Germany}
\author{Carsten Timm}
\email{carsten.timm@tu-dresden.de}
\affiliation{Institute of Theoretical Physics, Technische Universit\"at
Dresden, 01062 Dresden, Germany}

\date{\today}

\begin{abstract}
The nontrivial topology of the electronic structure of iron pnictides can lead
to the appearance of surface states. We study such states in various
strip geometries with a focus on the superconducting phase. In the presence
of unconventional superconducting pairing with $s_\pm$-wave gap structure, the
topological states are quite robust and partly remain in the
superconducting gap. Furthermore, Andreev bound states appear, which coexist
with the topological states for small superconducting gaps and merge with
them for larger gap values. The bulk and surface dispersions are obtained
from exact diagonalization for two-orbital and five-orbital models in
strip geometries.
\end{abstract}


\maketitle

\section{Introduction}

In the past few years, the iron pnictides\cite{Joh10,DHD12} on the one
hand and topological properties of matter\cite{HaK10,QiZ11} on the other have
been two of the most active fields in condensed-matter physics. Iron pnictides
feature unconventional multiband superconductivity with high transition
temperatures competing with itinerant antiferromagnetism. Topology is of
particular interest for condensed matter since nontrivial topological
properties of the band structure in the bulk are related to the existence of
surface or edge states. Previously, we have predicted that iron pnictides in
the paramagnetic and antiferromagnetic states can have surface states of
topological origin at (100) surfaces.\cite{LaT13} Here, we investigate the
surface states in the superconducting phase. We focus on the interplay of
surface states of topological origin with Andreev bound states enabled by
unconventional superconductivity.

The surface states in the paramagnetic phase and in the antiferromagnetic
spin-density-wave (SDW) phase result from winding of the
momentum-dependent Hamiltonian in orbital space, in particular
with respect to the iron $d_{xz}$ and $d_{yz}$ orbitals, noted already by Ran
\textit{et al.}\cite{RWZ09} The surface states
are of topological origin in the sense that the model Hamiltonian can be
deformed, without closing the gap existing in certain ranges of surface
momenta, into
one that is topologically nontrivial and has flat bands of surface states at
the Fermi energy. Reversing the deformation, the topological protection of
these surface states is lost, but they evolve \emph{continuously} as a function
of the deformation. Thus the surface bands become dispersive and generally move
away from the Fermi energy but are not destroyed until they merge with the
continuum of bulk bands.\cite{LaT13} The same type of argument can
explain the edge states at graphene zigzag edges, which form nearly but not
quite flat bands.\cite{CGP09,KOQ11} Since we are using two-dimensional
models, the surface states of slabs emerge as edge states of strips. Using
two-dimensional models corresponds to neglecting the dispersion in the $k_z$
direction. If we took the $k_z$ dispersion of the bulk bands into account,
the surface states would survive but also become dispersive.

Many iron pnictides show superconductivity in the vicinity of, or even
coexisting with,
an\-ti\-fer\-ro\-mag\-ne\-tism.\cite{TBN08,LKK09,PTK09,GAB09,LuC10,WLP11,ACG11}
It is therefore of interest how the surface states are modified when a
superconducting gap opens. The superconducting order parameter of the 1111
family of iron pnictides is thought to be of $s_\pm$-wave form, i.e., it has
opposite sign on the electron-like and the hole-like Fermi
pockets.\cite{HKM11,Chu12} The gap does not have nodes on the Fermi
surface. Andreev bound states have been studied for a simple two-band model
by Onari and Tanaka\cite{OnT09} and, within a quasiclassical approximation, by
Nagai \textit{et al.}\cite{NaH09} The latter group has extended their study to a
five-orbital model.\cite{NHM10} Huang and Lin\cite{HuL10} consider Anreev bound
states for a two-orbital model. We here find Andreev bound states inside the
superconducting gap that coexist with the topological surface states for small
gap magnitudes and merge with them at larger gap values. We also present
additional results for the paramagnetic and SDW phases, for strip orientations
not considered in Ref.\ \onlinecite{LaT13}. We will employ a simple
two-orbital model and a more realistic five-orbital
model.\cite{LaT13,RWZ09,KOA08}

The remainder of this paper is organized as follows. In Sec.\
\ref{sec:model}, we introduce the two-orbital and five-orbital models used in
our study. We then discuss the mean-field approximations for the SDW and
superconducting phases and the exact diagonalization for strip geometries.
In Sec.\ \ref{sec:res}, we present numerical results for the dispersion of
strips in the superconducting state, compare them to the paramagnetic and
antiferromagnetic states, and discuss the origin of the different
types of surface states. In Sec.\ \ref{sec:con} we summarize the results
and draw conclusions.

\section{Models and method}
\label{sec:model}

The two-orbital model of Ran \textit{et al.}\cite{RWZ09} is
formulated for a
two-dimensional iron square lattice and involves only the
$3d_{XZ}$ and $3d_{YZ}$ orbitals in a single-iron unit cell. The $X$ and
$Y$ axes are rotated by $45^\circ$ relative to the $x$ and $y$ axes of the
lattice. The noninteracting Hamiltonian used to model the
paramagnetic phase reads
$H_0 = \sum_{\mathbf{k}\sigma} \sum_{a,b=1}^2 \mathcal{H}_{ab}^0(\mathbf{k})\,
d_{\mathbf{k}a\sigma}^\dagger d_{\mathbf{k}b\sigma}$
with the $2\times 2$ matrix\cite{RWZ09}
\begin{eqnarray}
\lefteqn{ \mathcal{H}^0(\mathbf{k}) = 2t_1(\cos k_x - \cos k_y)\, \tau^1
} \nonumber \\
&& {}- 2(t_2 - t_2') \sin k_x \sin k_y\, \tau^3 \nonumber\\
&& {}+ [2(t_2+t_2') \cos k_x \cos k_y
  + 2t_1'(\cos k_x + \cos k_y)]\, \tau^0 \quad
\label{eq:2_orb_hamilton_matrix}
\end{eqnarray}
in orbital space.
Here, $\tau^1$, $\tau^2$, $\tau^3$ are Pauli matrices, $\tau^0$ is the unit
matrix, the index $1$ corresponds to $3d_{XZ}$, and the index $2$
to $3d_{YZ}$. The hopping parameters are chosen to be
$t_1=0.30\,\mathrm{eV}$, $t_1'=0.06\,\mathrm{eV}$, $t_2=0.51\,\mathrm{eV}$,
and $t_2'=0.09\,\mathrm{eV}$.\cite{RWZ09} Note that the band structure of
this model features quadratic band touching points in the center and at the
corners of the Brillouin zone (BZ).

In order to model the antiferromagnetic phase, we use the interacting
Hamiltonian $H=H_0+H_I$ with\cite{RWZ09}
\begin{eqnarray}
H_I &=& \frac{U}{2} \sum_\mathbf{j} (\hat{n}_{\mathbf{j}1}^2
+ \hat{n}_{\mathbf{j}2}^2) + (U-2J) \sum_\mathbf{j}
\hat{n}_{\mathbf{j}1}\, \hat{n}_{\mathbf{j}2}
\nonumber\\
&& {}+ J \sum_\mathbf{j} \sum_{\sigma \sigma'} d_{\mathbf{j}1\sigma}^\dagger
  d_{\mathbf{j}2\sigma'}^\dagger
  d_{\mathbf{j}1\sigma'}^{} d_{\mathbf{j}2\sigma}^{} \nonumber\\
&& {}+ J \sum_\mathbf{j} (d_{\mathbf{j}1\uparrow}^\dagger
  d_{\mathbf{j}1\downarrow}^\dagger
  d_{\mathbf{j}2\downarrow}^{} d_{\mathbf{j}2\uparrow}^{} + \textrm{H.c.}),
\label{eq:2_orb_HI}
\end{eqnarray}
where $\hat{n}_{\mathbf{j}a} \equiv \sum_\sigma d_{\mathbf{j}a\sigma}^\dagger
d_{\mathbf{j}a\sigma}^{}$. For the interaction parameters, we take
$U=1.20\,\textrm{eV}$ and $J=0.12\,\textrm{eV}$.~\cite{RWZ09} A mean-field
decoupling of the form~\cite{LaT13}
$\langle d_{\mathbf{j}a\sigma}^\dagger d_{\mathbf{j}b\sigma}\rangle
= n_{ab} + \frac{\sigma}{2}\, (-1)^{j_x}\, m_{ab}$,
where we have assumed an SDW ordering vector $\mathbf{Q}=(\pi,0)$ and spins
pointing along the $S_z$ axis, then leads to the mean-field
Hamiltonian
\begin{equation}
H_\mathrm{MF} = H_0 + \sum_\mathbf{j} \sum_{a,b} (-1)^{j_x} M_{ab}\,
(d_{\mathbf{j}a\uparrow}^\dagger d_{\mathbf{j}b\uparrow}^{} -
d_{\mathbf{j}a\downarrow}^\dagger d_{\mathbf{j}b\downarrow}^{}),
\label{eq:2_orb_mean_field_bulk}
\end{equation}
with $M_{11} = -(Um_{11}+Jm_{22})/2$, $M_{22}=-(Um_{22}+Jm_{11})/2$, and
$M_{12} = M_{21} = -Jm_{12} = -Jm_{21}$.\cite{LaT13}
The mean-field coefficients $m_{ab}$ are calculated self-con\-sis\-tent\-ly
assuming half filling.

\begin{figure}[t]\centering
\subfloat{\includegraphics[width=0.48\linewidth]
{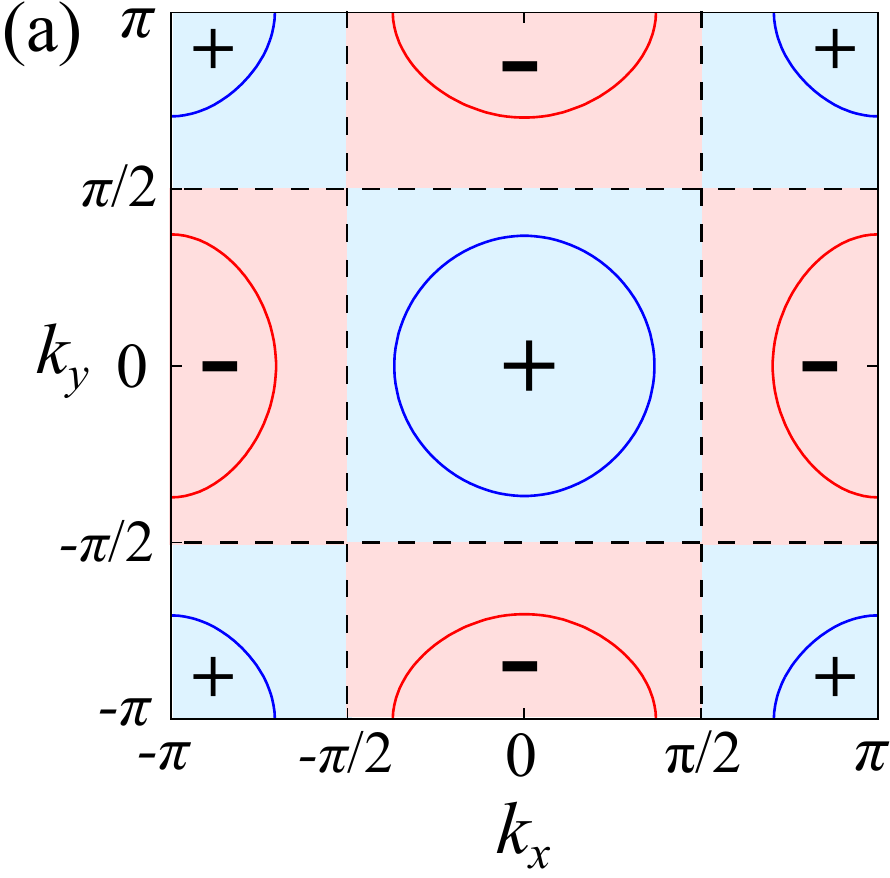}}
\hfill
\subfloat{\includegraphics[width=0.48\linewidth]
{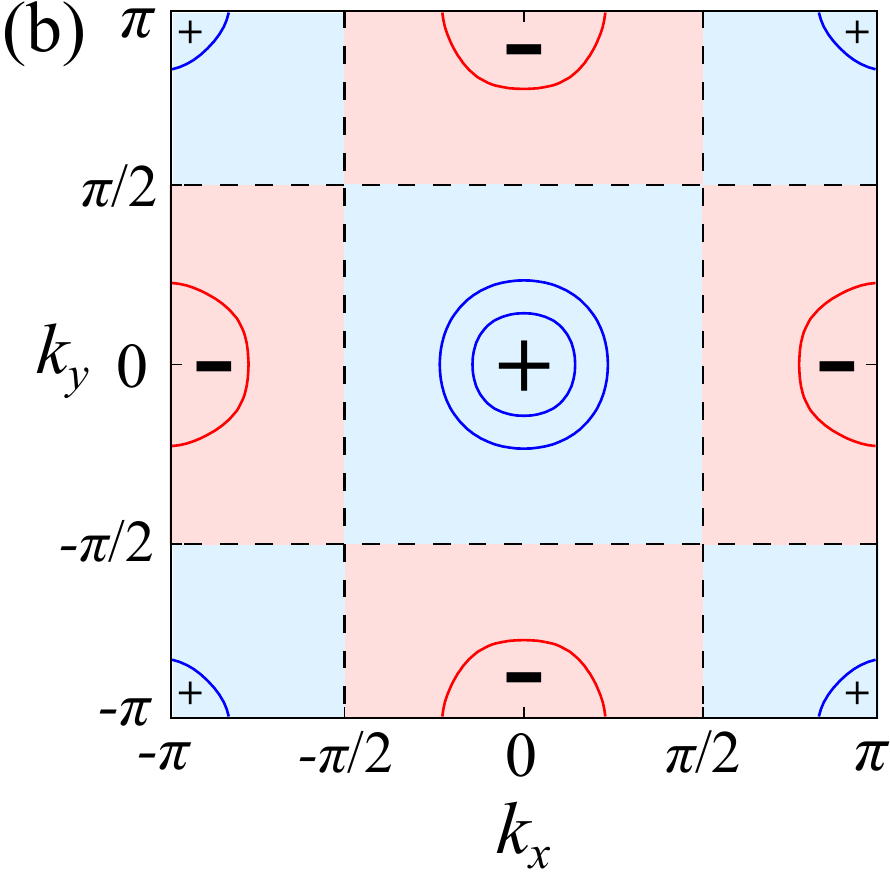}}
\caption[Illustration of the $s_\pm$-wave gap structure]
{(Color online) Sign structure of the superconducting gap function with
$s_\pm$-wave structure.
The red (blue) areas denote regions in the BZ where the sign of $\cos k_x
\cos k_y$ is negative (positive). The dashed lines show the nodes of this
function. In addition, the Fermi surfaces of (a) the two-orbital model
and (b) the five-orbital model are plotted.}
\label{fig:sc_sign_structure}
\end{figure}

For the superconducting phase, we employ the BCS mean-field
Hamiltonian
\begin{eqnarray}
\lefteqn{ H_\mathrm{BCS} = \sum_{\mathbf{k}\sigma}\sum_{ab}
  [\mathcal{H}^0_{ab}(\mathbf{k})-\delta_{ab}
  \mu]\, d_{\mathbf{k}a\sigma}^\dagger d_{\mathbf{k}b\sigma}^{} }
  \nonumber\\
&& {}- \sum_\mathbf{k}\sum_{ab} \big[ \Delta_{ab}(\mathbf{k})\,
  d_{\mathbf{k}a\uparrow}^\dagger d_{-\mathbf{k},b,\downarrow}^\dagger
  + \Delta_{ab}^* (\mathbf{k})\,
  d_{-\mathbf{k},b,\downarrow}^{} d_{\mathbf{k}a\uparrow}^{} \big] ,\quad
\label{eq:sc_BCS_Hamiltonian_bulk}
\end{eqnarray}
where $\mu$ denotes the chemical potential at half filling.
For the superconducting gap function we use
$\Delta_{ab}(\mathbf{k}) = \Delta\,\delta_{ab}$ for
conventional $s_{++}$-wave pairing and
$\Delta_{ab}(\mathbf{k}) = \Delta\,\delta_{ab}\, \cos k_x \cos k_y$ for the
unconventional $s_\pm$-wave pairing likely realized in 1111
iron pnictides.\cite{HKM11,Chu12} The sign structure of the gap function with
$s_\pm$-wave gap structure is illustrated in Fig.~\ref{fig:sc_sign_structure}
along with typical Fermi surfaces for the two-orbital and  five-orbital
models. Note that there is a sign
change of the superconducting gap between electron and hole Fermi
pockets. The corresponding Bogoliubov-de Gennes (BdG) Hamiltonian for our
model is a $4\times 4$ matrix, which reads\cite{HuL10}
\begin{equation}
\mathcal{H}_\mathrm{BdG} =
\begin{pmatrix}
\mathcal{H}^0(\mathbf{k})- \mu \tau^0 & -\Delta(\mathbf{k}) \\
-\Delta^\dagger(\mathbf{k}) & -\mathcal{H}^0(\mathbf{-k})+ \mu \tau^0
\end{pmatrix}
\end{equation}
with respect to the basis
$\lbrace d_{\mathbf{k}1\uparrow}^\dagger,\, d_{\mathbf{k}2\uparrow}^\dagger,\,
d_{-\mathbf{k},1,\downarrow}^{},\,d_{-\mathbf{k},2,\downarrow}^{}\rbrace$.
The two-orbital model\cite{RWZ09} used here and in Ref.\ \onlinecite{HuL10}
is different from the model employed for the study of Andreev bound
states in Refs.\ \onlinecite{OnT09,NaH09}. The latter has only one electron and
one hole Fermi pocket each and is effectively rotated by $45^\circ$ compared
to our model.

The more realistic five-orbital model of Kuroki \textit{et
al.}\cite{KOA08} includes
all hopping amplitudes larger than $10\,\mathrm{meV}$ up to fifth neighbors.
Along with the onsite energies, they are obtained from density-functional
calculations for LaFeAsO and are tabulated in Ref.~\onlinecite{KOA08}.
Moreover, the orbital indices now assume values $a,b=1,\ldots,5$ corresponding
to $3d_{3Z^2-R^2}$, $3d_{XZ}$, $3d_{YZ}$, $3d_{X^2-Y^2}$, $3d_{XY}$,
respectively. Note that the band structure of the five-orbital model exhibits,
besides quadratic band touching points, also Dirac points.

The interaction Hamiltonian for the antiferromagnetic phase is basically the
same as for the two-orbital model, except that the interorbital terms
in Eq.~\eqref{eq:2_orb_HI} now become sums over all pairs of five
orbitals. For the interaction parameters we take
$U=1.0\,\mathrm{eV}$ and $J=0.2\,\mathrm{eV}$.\cite{RWZ09} A decoupling
as above then yields a mean-field Hamiltonian analogous to
Eq.~\eqref{eq:2_orb_mean_field_bulk}. The corresponding coefficients $M_{ab}$
are given in Ref.~\onlinecite{LaT13}. The mean-field parameters $m_{ab}$ are
calculated self-consistently, assuming $6$ electrons per iron, corresponding to
zero doping. The BCS Hamiltonian for the superconducting phase is
analogous to the two-orbital case, taking the larger number of orbitals
into account.

We are interested in edge states of strips described by the two
models. Specifically, we investigate strips of width $W$ with (10),
(01), and (11) edges. We assume that the SDW and superconducting strips are
described by the same uniform order parameters $m_{ab}$ and $\Delta$ as the
bulk systems. We briefly return to this point in Sec.~\ref{sec:con}.

For a strip with (10) edges, $k_y$ is still a good quantum number since
the strip is extended along the $y$ axis. Therefore, we carry out a Fourier
transformation in the $y$ direction,
$d_{\mathbf{j}a}^\dagger = N_y^{-1/2}\sum_{k_y} e^{-ik_y j_y}\,
d_{j_x k_y a}^\dagger$, giving a block-diagonal Hamiltonian with blocks
enumerated by $k_y$. The dimension of the blocks is a multiple of $W$. The
energy bands of the (10) strip are then obtained by exact diagonalization
of these blocks. For strips with (01) edges, we simply interchange the
roles of $x$ and $y$.

\begin{figure}[t]\centering
\includegraphics[width=1.0\linewidth]{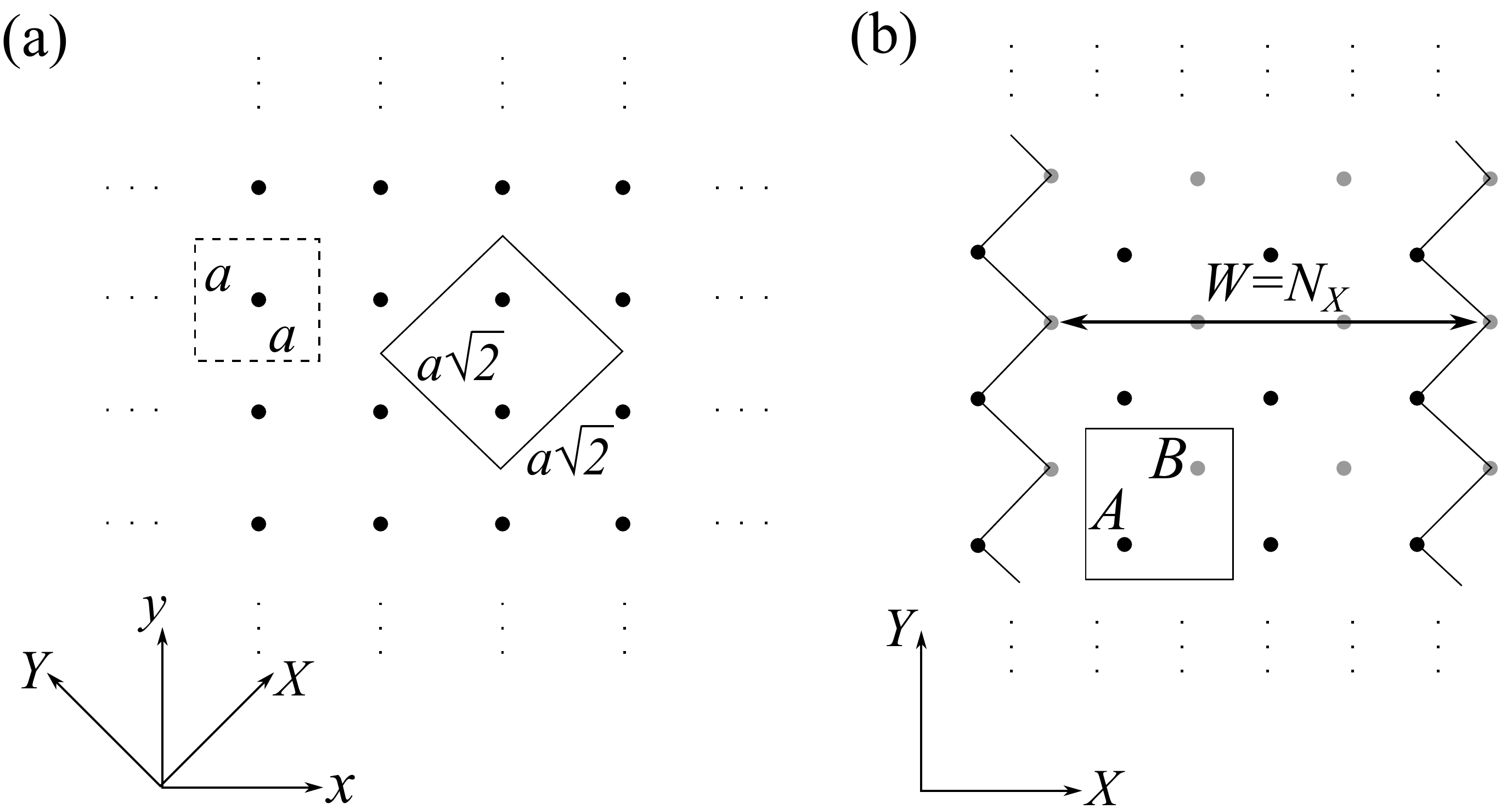}
\caption[New unit cell and (11) edges]
{New unit cell and (11) edges: (a) Comparison of the old unit cell (dashed
square) and the new unit cell (solid square). The new unit cell is rotated by
$45^\circ$ and its sides are stretched by a factor of $\sqrt{2}$. $a$ is the
lattice constant. (b) Strip with two (11) edges in the rotated coordinate
system. Black (grey) dots belong to sublattice $A$ ($B$). The strip width is
$W=N_X$. The new unit cell is depicted by a solid square.}
\label{fig:unit_cell_edge_11}
\end{figure}

Strips with (11) edges require a different treatment since the
edges cut diagonally through the lattice. It is convenient to use a
unit cell with two sides parallel to the (11) edges, see
Fig.~\ref{fig:unit_cell_edge_11}. The new unit cell contains two iron
sites. For this reason, we can describe the lattice in terms of
two quadratic sublattices $A$ and $B$. We further introduce a new
coordinate system, whose axes, denoted by $X$ and $Y$, are rotated by $45^\circ$
with respect to the old coordinate system and are thus aligned with
the sides of the new unit cell. After representing $H_0$, $H_\mathrm{MF}$,
and $H_\mathrm{BCS}$ in the new coordinates, we perform a Fourier
transformation in the $Y$ direction,
\begin{eqnarray}
d_{j_X j_Y a A}^\dagger
&=& \frac{1}{\sqrt{N_Y}} \sum_{k_Y}
e^{-i\sqrt{2}\,k_Y j_Y}\, d_{j_X k_Y a A}^\dagger,
\label{eq:2_orb_fourier_transforms_11_1}\\
d_{j_X j_Y a B}^\dagger
&=& \frac{1}{\sqrt{N_Y}} \sum_{k_Y}
e^{-i\frac{k_Y}{\sqrt{2}}}\, e^{-i\sqrt{2}\,k_Y j_Y}\, d_{j_X k_Y a B}^\dagger,
\label{eq:2_orb_fourier_transforms_11_2}
\end{eqnarray}
where $k_Y\in(-\pi/\sqrt{2},\pi/\sqrt{2}]$ and $N_Y$ is the number of unit
cells along the $Y$ axis. This leads to a block Hamiltonian, which we
diagonalize to obtain the energy bands.

\section{Results and discussion}
\label{sec:res}

\subsection{Strips with (10) or (01) edges}
\label{sub:1001}

\subsubsection*{Paramagnetic and antiferromagnetic phase}

Our results for strips with (10) edges in the paramagnetic and in the
antiferromagnetic phase, obtained in Ref.~\onlinecite{LaT13}, are briefly
summarized in the following. In the two-orbital model,
four bands of edge states are present in the paramagnetic phase. They
are exactly degenerate in pairs due to SU(2) spin-rotation symmetry. The two
pairs are bonding and anti-bonding combinations of states localized at the two
edges and become degenerate in the limit of a broad strip. In the five-orbital
model, two such groups of four nearly degenerate bands appear.
As noted in the introduction, the existence of surface states can be
understood from an argument based on a continuous deformation, which
does not close the gap, of the Hamiltonian into a topologically nontrivial
one. Upon turning on SDW order with ordering vector $\mathbf{Q}=(\pi,0)$,
the asymptotically degenerate bundles of bands split
due to the coupling of the spin of the
electrons localized at the surface to the SDW order parameter, which is
uniform along the (10) edges. However, they remain
exactly degenerate in pairs since the corresponding mean-field Hamiltonian is
still invariant under combined spin rotation by $\pi$ about the $x$
axis and spatial reflection $x\rightarrow -x$.

For the antiferromagnetic phase with ordering vector $(\pi,0)$,
the (01) edge is not equivalent to the (10) edge. For the (01) edge,
the magnetic unit cell is doubled in the $x$ direction and thus the
one-dimensional (1D) edge BZ is halved. As a consequence, the number of surface
bands doubles, due to the folding of the spectrum, and the resulting degeneracy
at the boundaries of the magnetic BZ is lifted by the SDW. Unlike for the
(10) edge, the original four-fold degeneracy of the surface bands for
$W\to\infty$ remains intact (not shown). This is because the
magnetization at the (01) edges is staggered so that states of opposite spin
localized at the same edge are not split. These observations hold for both
models.

\subsubsection*{Superconducting phase}

\begin{figure}[t]\centering
\subfloat{\includegraphics[width=0.495\linewidth]
{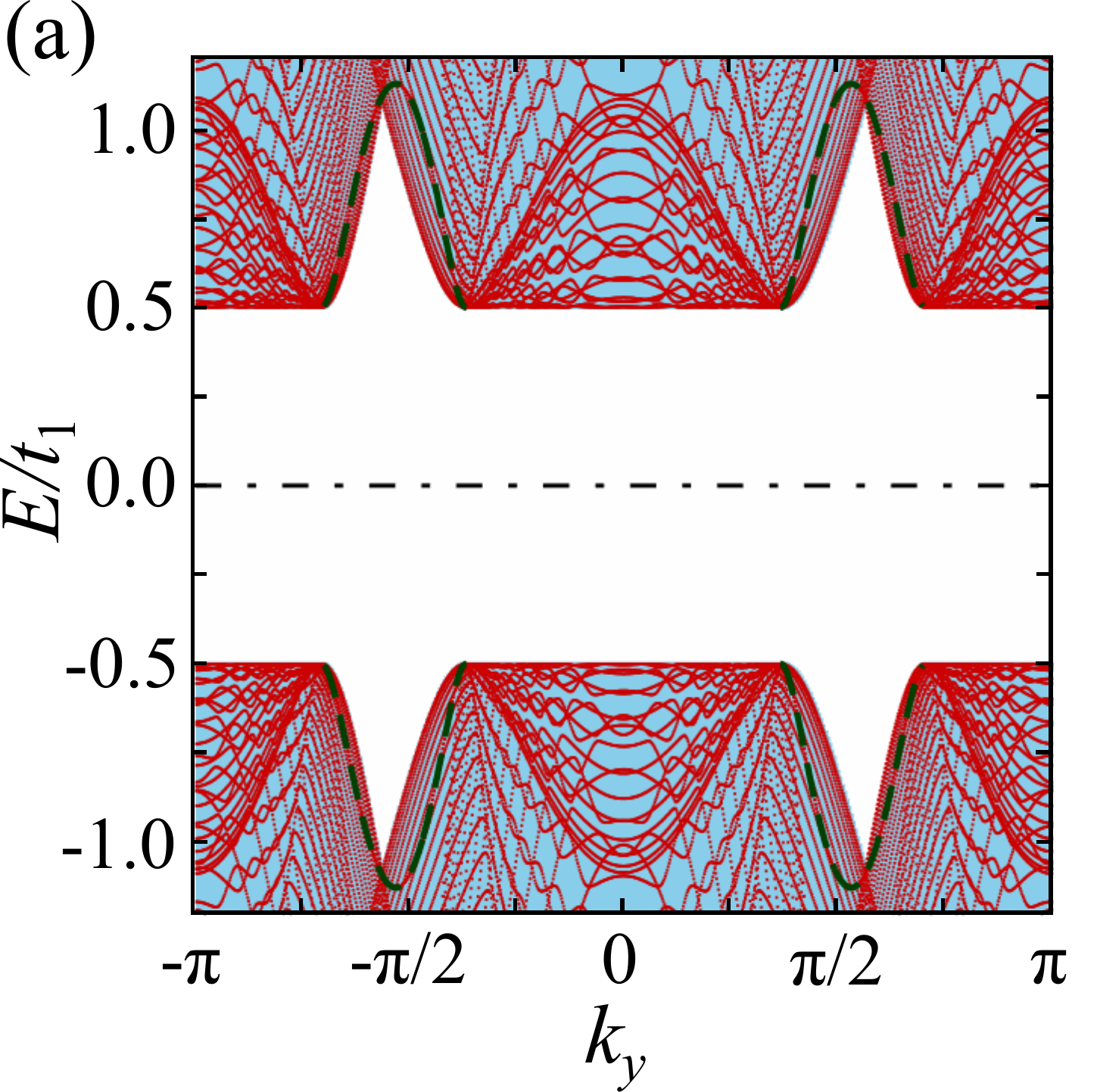}}
\hfill
\subfloat{\includegraphics[width=0.495\linewidth]
{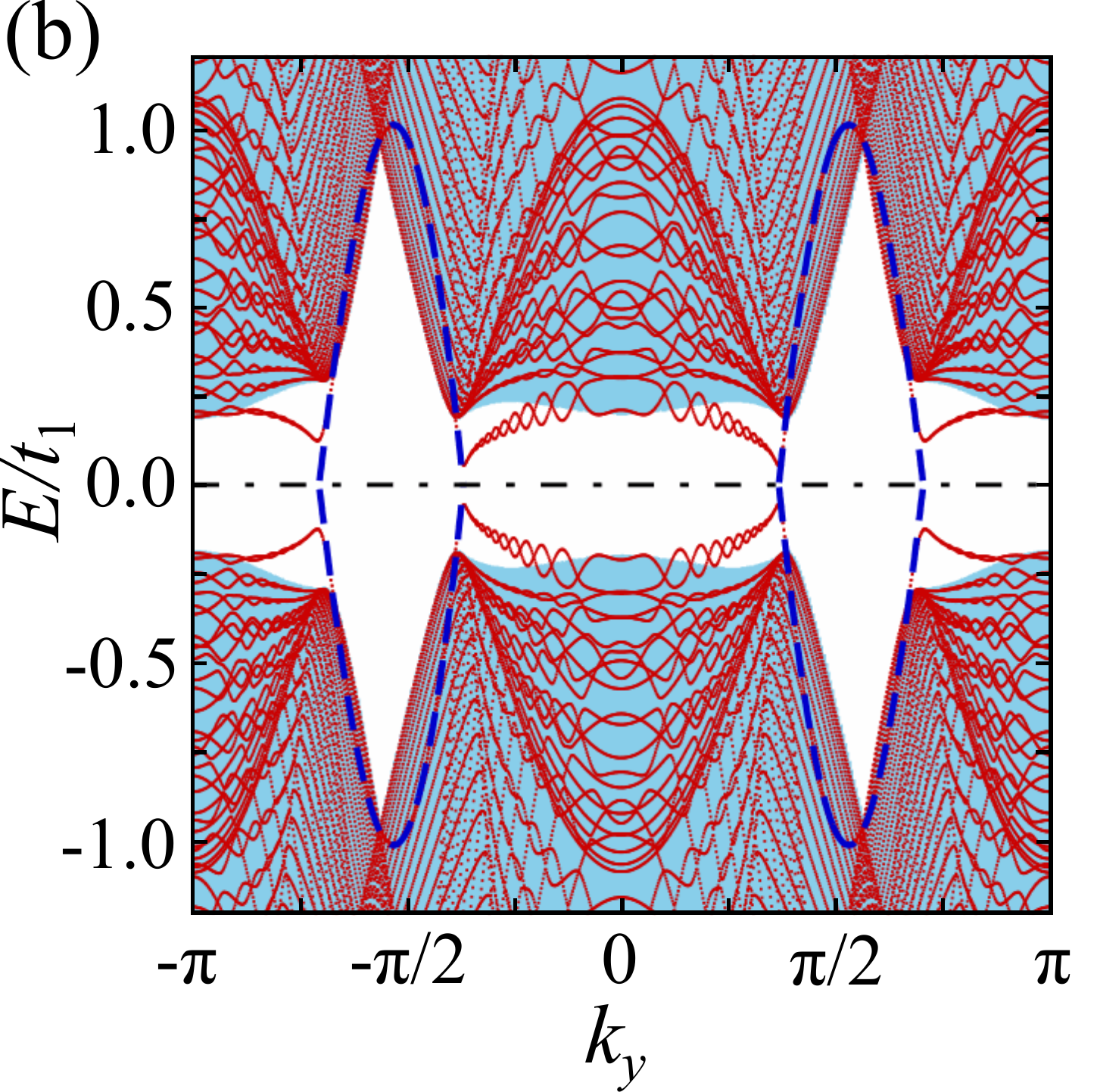}}\\
\vspace{-1em}
\subfloat{\includegraphics[width=0.495\linewidth]
{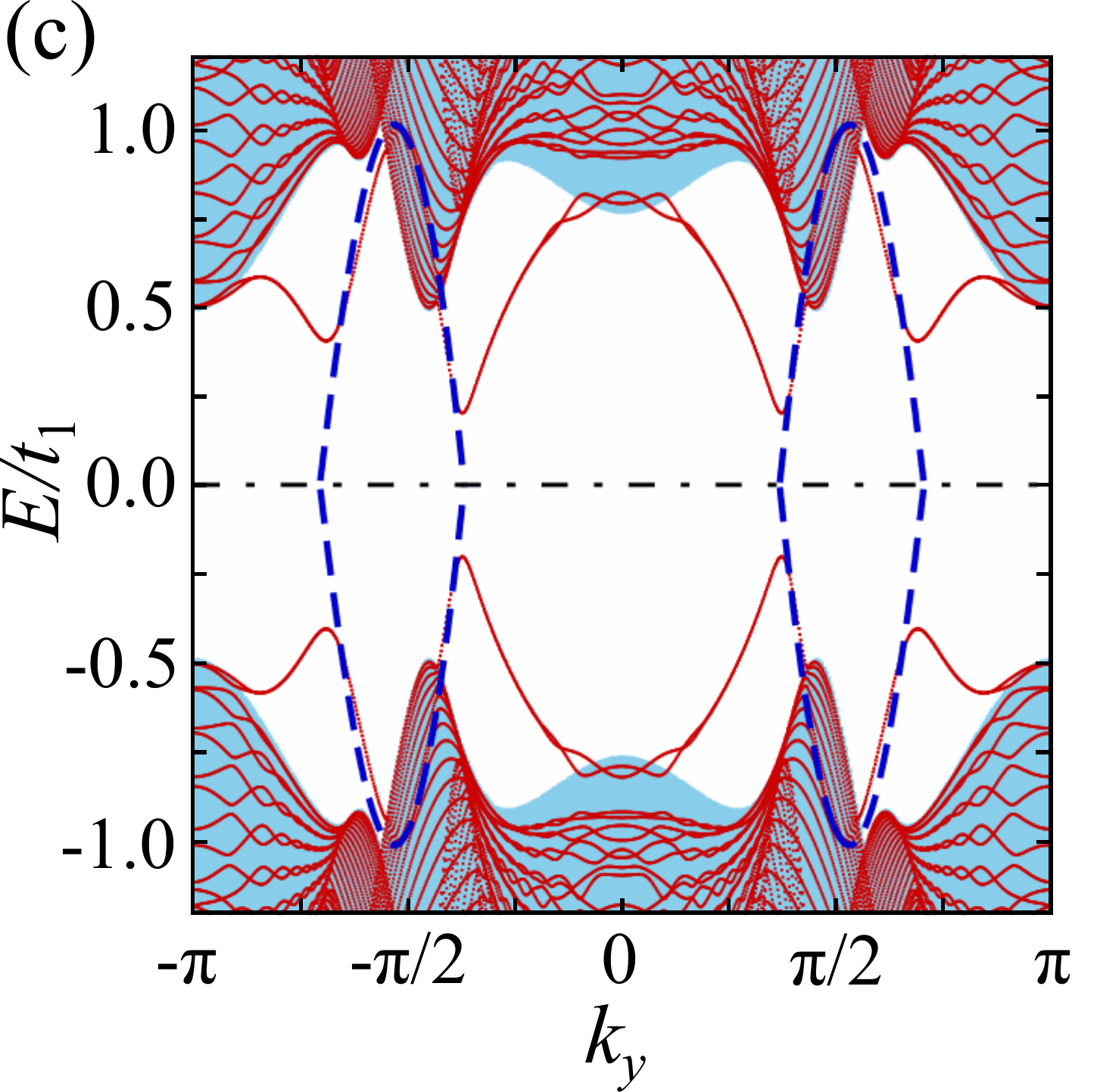}}
\hfill
\subfloat{\includegraphics[width=0.495\linewidth]
{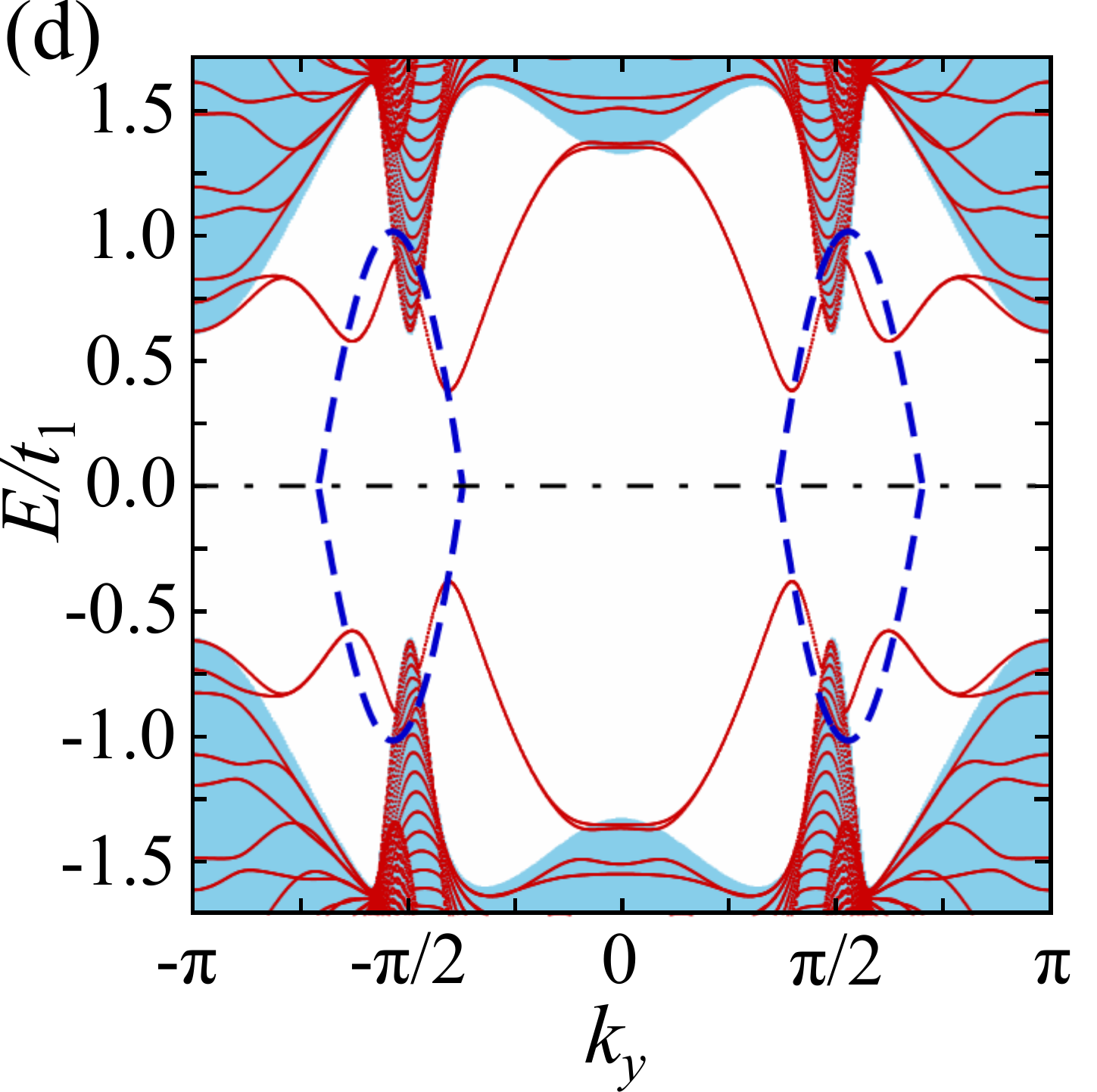}}
\caption[Quasi-particle spectra for the superconducting (10)
strip in the two-orbital model]
{(Color online) Quasi-particle spectra in the superconducting phase
for the two-orbital model.
Bands of a (10) strip of width $W=40$ (red) are compared to the bulk bands
projected onto the 1D BZ for the strip (blue). Only the low-energy part of
the spectra is shown.
(a)~$\Delta=0.5$ for $s_{++}$-wave pairing. For comparison,
the topological surface bands, modified according to $\xi\to\pm\sqrt{\xi^2
+ |\Delta|^2}$, are also plotted (dashed green lines).
(b)~$\Delta=0.5$ for $s_{\pm}$-wave pairing,
(c)~$\Delta=2.0$ for $s_{\pm}$-wave pairing,
(d)~$\Delta=6.0$ for $s_{\pm}$-wave pairing.
In panels (b)--(d), the dashed dark blue lines denote topological
surface bands of the normal
state.}
\label{fig:2_orb_bands_sc_10}
\end{figure}

We now consider the superconducting phase, starting with the two-orbital
model. Quasi-particle spectra of the superconducting (10) strip are shown
for $s_{++}$-wave pairing in Fig.\ \ref{fig:2_orb_bands_sc_10}(a) and for
$s_\pm$-wave pairing in Figs.\ \ref{fig:2_orb_bands_sc_10}(b)--(d).
In all cases, the bands for the strip are compared to the
bulk bands projected onto the 1D BZ for the strip. Large
values of the gap $\Delta$ have been considered to more clearly exhibit the
effects of interest. The spectra show the typical doubling and particle-hole
symmetry induced by the BdG description.

In the case of $s_{++}$-wave pairing, a full gap opens
without any edge states inside the gap. The bulk states are pushed out of
the gap according to $\xi\to\pm\sqrt{\xi^2 + |\Delta|^2}$. Interestingly,
the surface bands are modified in the same way, as emphasized by the
dashed green
lines in Fig.~\ref{fig:2_orb_bands_sc_10}(a). We can understand this by noting
that the $s_{++}$-wave pairing interaction is purely local and is therefore
not affected by the introduction of edges. Hence, one would indeed expect
$s_{++}$-wave pairing to induce similar gaps for bulk and surface states.
In this process, the edge bands from the normal state (dashed green
lines) become resonant with bulk states, which destroys their localization at
the edges. Moreover, there are no Andreev bound states. This is expected
since for Andreev bound states to appear Andreev reflection involving gaps
of opposite sign has to be possible.\cite{KaT00}

Let us now discuss the realistic case of $s_\pm$-wave superconductivity.
First, Figs.~\ref{fig:2_orb_bands_sc_10}(b)--(d) show that the bulk gap is
no longer constant in the BZ. Moreover, there are states inside the bulk
gap. In contrast to the $s_{++}$-wave case,
the topological surface bands from the normal state are not pushed away.
In fact, they coincide closely with the normal-state bands and their charge
conjugates, as indicated by the dashed dark blue lines. Although they are still
mostly hidden in the bulk continuum in Fig.\
\ref{fig:2_orb_bands_sc_10}(b), parts of them become visible within the
gap. Near zero energy, we observe a gap for the
surface bands, which is much smaller than the bulk gap.
Furthermore, we find additional edge bands in ranges of $k_y$ without edge
states in the normal phase, but connected to them. They merge with the bulk
continuum at $k_y=0$ and $k_y=\pi$. On the whole, there is a pair of surface
bands for $E>0$, doubled at $E<0$. Within the pairs, we find bonding and
anti-bonding states whose energy difference is exponentially small for
large width.

It is reasonable that $s_\pm$-wave and $s_{++}$-wave pairing differently
affect the surface states resulting from the normal phase. The
$s_\pm$-wave pairing interaction is not
local but connects next-nearest-neighbor sites. Thus, the
interaction is cut off at the edges so that it affects edge states
less strongly than bulk states.

\begin{figure}[t]\centering
\subfloat{\includegraphics[width=0.48\linewidth]
{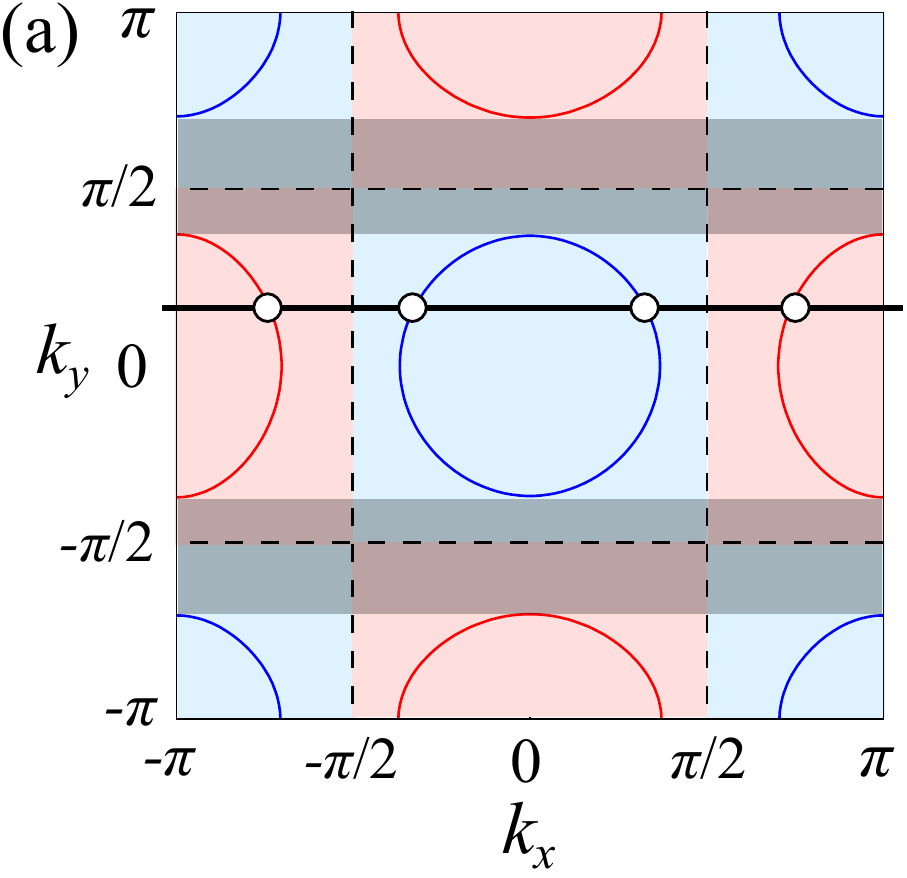}}
\hfill
\subfloat{\includegraphics[width=0.48\linewidth]
{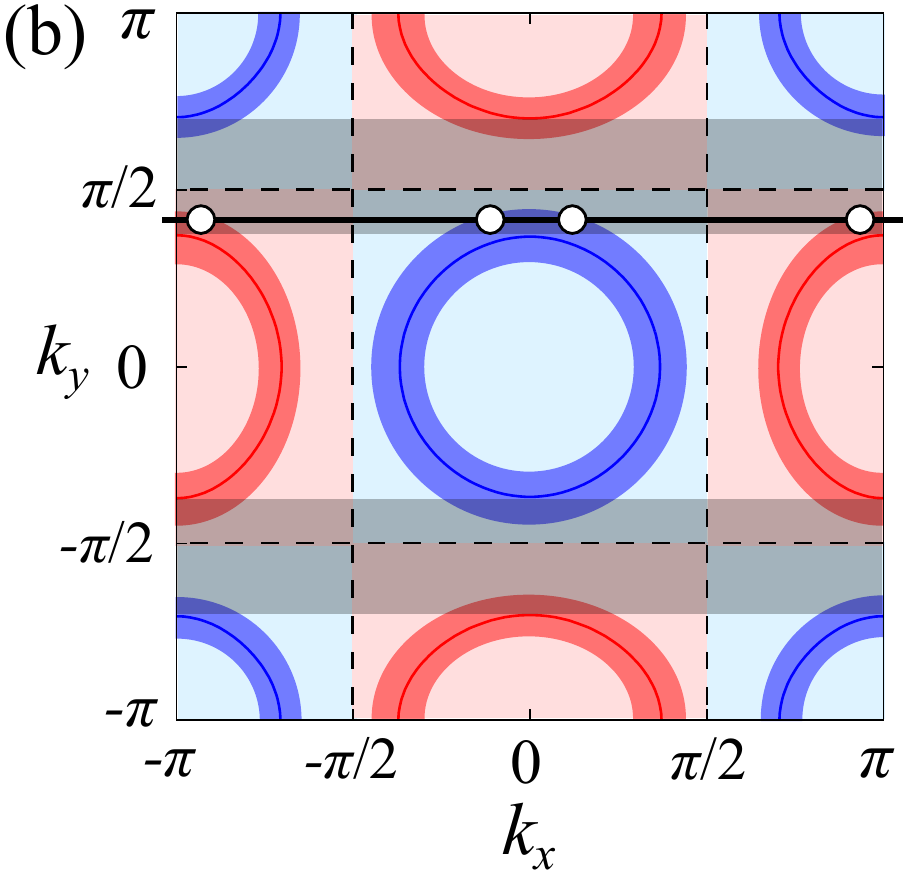}}
\caption[Scattering processes for the (10) strip]
{(Color online) Scattering processes in the two-orbital model for the (10)
strip. The shaded regions indicate $k_y$ values for which surface states of
topological origin exist in the normal, paramagnetic phase. The bold lines
are exemplary lines with constant $k_y$ for which sign-changing scattering
processes are possible.
Relevant states are indicated by white circles. (a) Small $\Delta$:
sign-changing processes can only occur outside of the shaded regions.
Topological states and Andreev bound states are separated. (b) Larger $\Delta$:
states in a broader region around the Fermi surfaces become relevant.
Topological states merge with Andreev bound states.}
\label{fig:sc_scattering_10}
\end{figure}

The additional surface bands can be explained as Andreev bound
states:\cite{KaT00,OnT09,NaH09}
The Fermi surface of the normal state along with the sign structure
of the gap function are illustrated in Fig.~\ref{fig:sc_scattering_10}.
The edge is parallel to the $y$ axis and, hence, $k_y$ is a constant of
motion during the scattering processes. Therefore, we have to consider
lines through the BZ with constant $k_y$ in
order to find the available states. Furthermore, for small gap amplitudes
$\Delta$, only states at the Fermi surfaces are relevant. From
Fig.~\ref{fig:sc_scattering_10}(a), we see that sign-changing scattering
processes are possible for all $k_y$ except where the line $k_y=\mathrm{const}$
does not cross a Fermi surface. But the latter is exactly the region where we
have found topological surface states.
In other words, topological states are only possible if the
$k_y=\mathrm{const}$ line corresponds to a gapped system with \emph{no}
states available at the Fermi energy, whereas Andreev states require a gapless
system where states at the Fermi level \emph{do} exist.
Hence, topological surface states
inherited from the normal phase and Andreev bound states coexist,
but their $k_y$ ranges do not overlap in the limit of small $\Delta$.

Note that the bands of topological states and of Andreev bound
states are connected. This is easy to understand: As discussed above, the
bands of topological surface states are less strongly affected by the nonlocal
$s_\pm$-wave pairing. On the other hand, the bands must be continuous in $k_y$
and thus cannot suddenly terminate. Consequently, in the superconducting state
additional states must appear in the gap that complete the bands of topological
states.

For larger $\Delta$, states from a broader range of $\mathbf{k}$
values in the vicinity of the normal-state Fermi surface are relevant
for Andreev scattering, as illustrated by
Fig.~\ref{fig:sc_scattering_10}(b). Consequently, the $k_y$ range for Andreev
bound states grows, as does the transition region between them and the
topological surfaces states. In Figs.~\ref{fig:2_orb_bands_sc_10}(b)--(d),
the effect of a growing gap amplitude is depicted. We observe that the gap
in the surface bands gets larger. Moreover, in the $k_y$ range for
topological surface
states, the edge bands lose their resemblance to the normal state (dashed
dark blue lines). This is
due to both the large gap amplitude, which now strongly affects also the
topological surface states, and the growing contribution of Andreev scattering.

The bands of Andreev bound states in Fig.\ \ref{fig:2_orb_bands_sc_10} are
similar to the ones found in Ref.\ \onlinecite{HuL10}. They are also
qualitatively similar to the bands in Ref.\ \onlinecite{OnT09} for the (11)
edge, which corresponds to our (10) edge due to the $45^\circ$ rotation of
the BZ. Surface states resulting from the normal phase are not addressed
in either work.

\begin{figure}[t]\centering
\subfloat{\includegraphics[width=0.495\linewidth]
{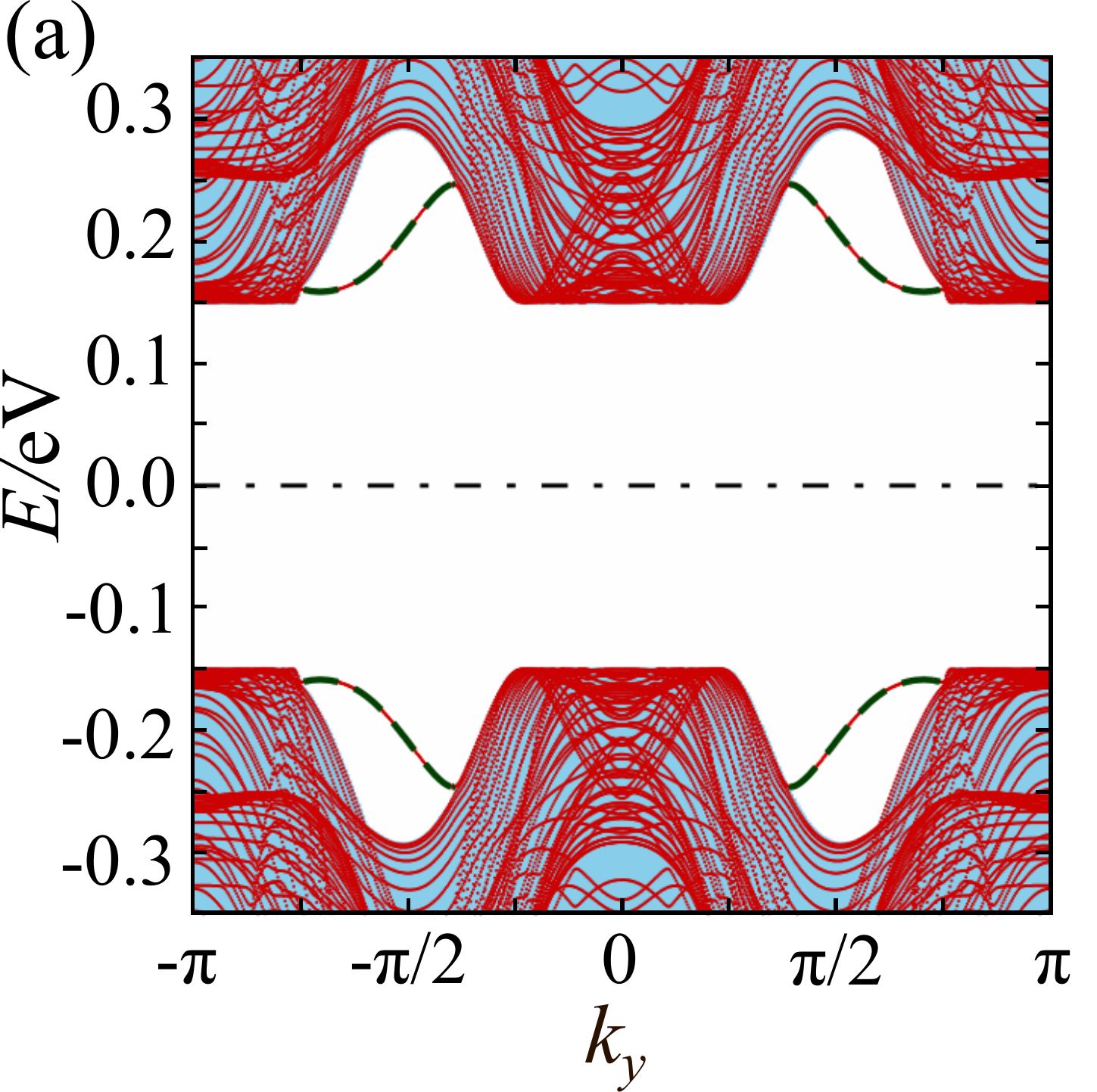}}
\hfill
\subfloat{\includegraphics[width=0.495\linewidth]
{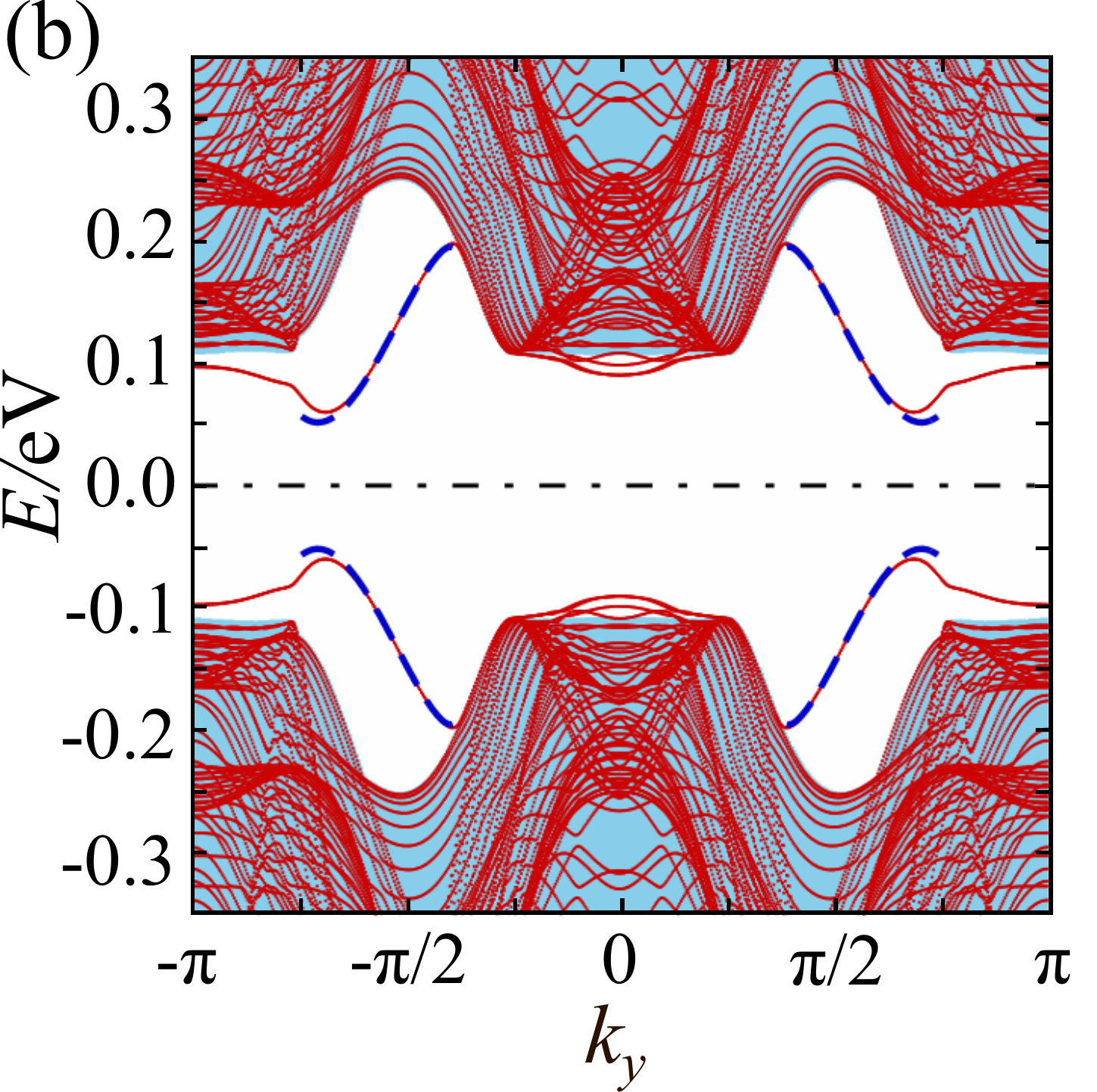}}\\
\vspace{-0.5em}
\subfloat{\includegraphics[width=0.495\linewidth]
{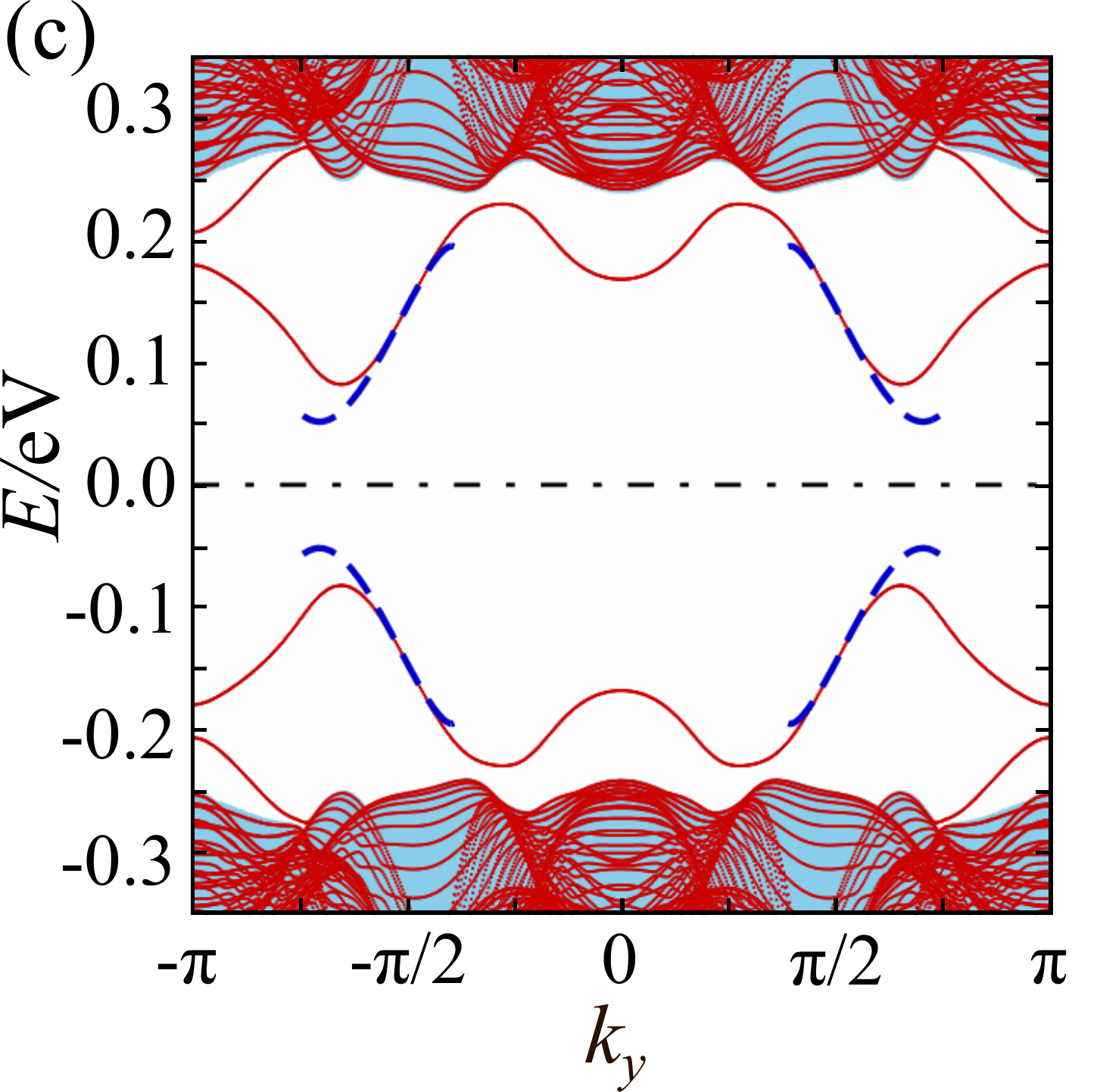}}
\caption[Quasi-particle spectra for the superconducting (10)
strip in the five-orbital\newline model]
{(Color online) Quasi-particle spectra in the superconducting phase for the
five-orbital model.
Bands of a (10) strip of width $W=40$ (red) are compared to the bulk bands
projected onto the 1D BZ for the strip (blue). Only the low-energy part of
the spectra is shown.
(a)~$\Delta=0.15$ ($s_{++}$ pairing). For comparison, some of
the topological surface bands modified according to $\xi\to\pm\sqrt{\xi^2
+ |\Delta|^2}$ are also plotted (dashed green lines).
(b)~$\Delta=0.15$ ($s_{\pm}$ pairing), (c)~$\Delta=0.4$ ($s_{\pm}$ pairing).
In panels (b) and (c), the dashed dark blue lines denote
topological surface bands of the
normal state.}
\label{fig:5_orb_bands_sc_10}
\end{figure}

We now turn to the five-orbital model. Quasi-particle spectra of
the superconducting (10) strip are shown for $s_{++}$-wave pairing in Fig.\
\ref{fig:5_orb_bands_sc_10}(a) and for $s_\pm$-wave pairing in Figs.\
\ref{fig:5_orb_bands_sc_10}(b), (c).
For $s_{++}$-wave pairing, we observe that bulk and surface states are
affected similarly by superconductivity, as for the two-orbital model. In
the five-orbital model, there are two bundles of surface bands in the normal,
paramagnetic phase.\cite{LaT13}
In the $s_{++}$-wave superconducting state, one of these bundles vanishes
in the bulk continuum. However, in contrast to the two-orbital model, the lower
bundle remains visible in the bulk gap. Its dispersion is well represented
by modifying the normal-state band according to $\xi\to\pm\sqrt{\xi^2
+ |\Delta|^2}$ (dashed green lines). Andreev bound states are not present
due to the absence of sign-changing scattering processes.

The case of $s_\pm$-wave superconductivity is illustrated in
Figs.~\ref{fig:5_orb_bands_sc_10}(b), (c). Like for the two-orbital model,
we see that the topological surface bands are hardly affected by
superconductivity, except close to the Fermi energy, where the bands are
pushed to higher energies. The effect is weaker than for the
two-orbital model because the normal-state surface bands do not lie close
to the Fermi energy. The higher-energy topological surface band is not visible
since it is resonant with the bulk continuum.
Like for the two-orbital model, we observe Andreev bound states
outside of the $k_y$ range for which we have found topological edge states in
the normal phase. The explanation is analogous to the two-orbital
model. For increasing gap amplitude, the
Andreev bound states merge with the lower-energy topological surface band,
which loses its resemblance to the normal state, as for the two-orbital model.
Eventually, the entire band separates from the bulk continuum,
as seen in Fig.~\ref{fig:5_orb_bands_sc_10}(c).
Along with this, a second band of Andreev bound states appears.

\subsection{Strips with (11) edges}

\subsubsection*{Paramagnetic and antiferromagnetic phase}

\begin{figure}[t]\centering
\includegraphics[width=0.95\linewidth]{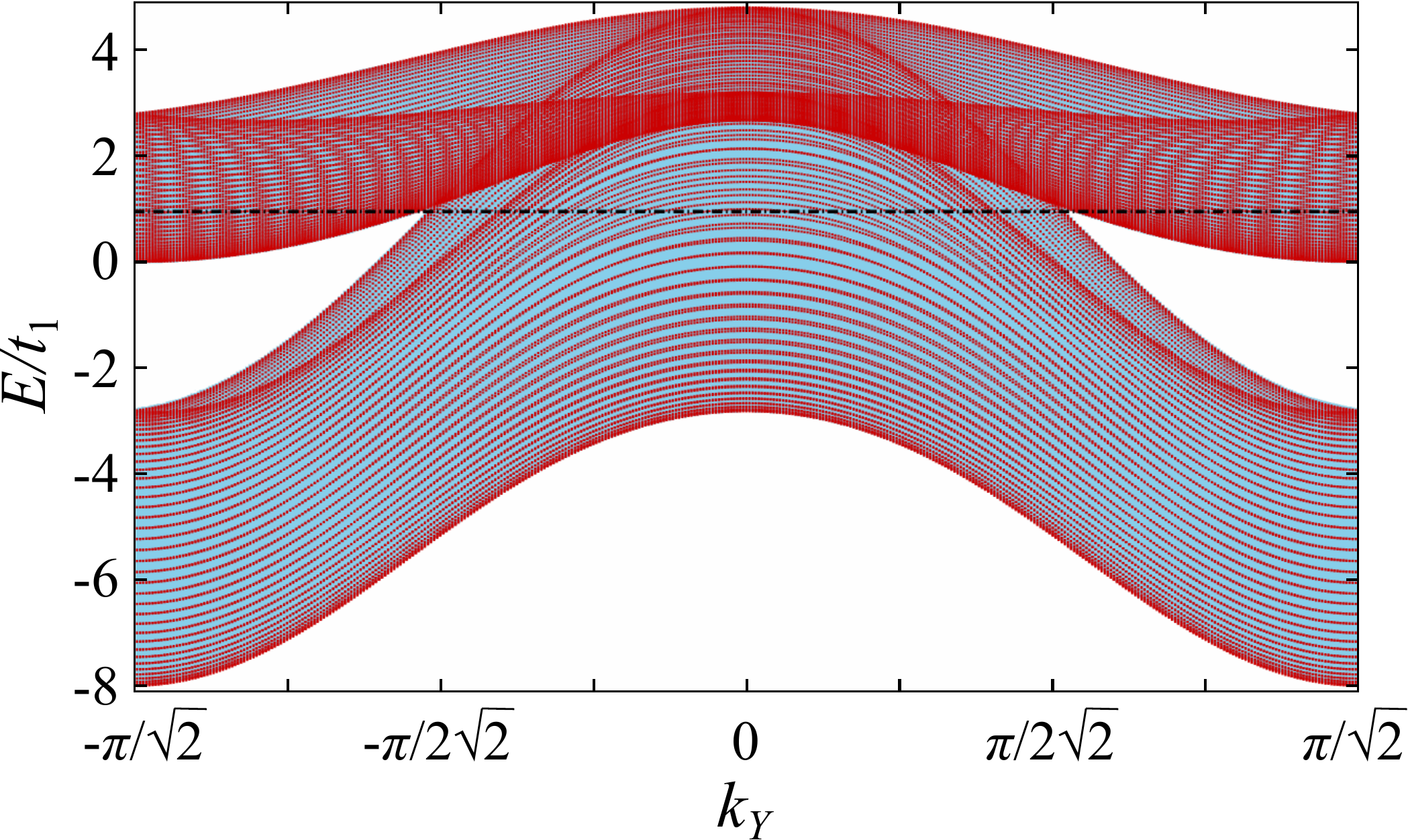}
\caption[Energy bands for the paramagnetic (11) strip in the two-orbital model]
{(Color online) Energy bands in the paramagnetic phase for the two-orbital
model. Bands of a (11) strip of width $W=40$ (red) are compared to the
bulk bands projected onto the 1D BZ for the strip (blue). The black dash-dotted
line denotes the Fermi energy at half filling.}
\label{fig:2_orb_bands_paramagnetic_11}
\end{figure}

For the (11) strip, we begin with the discussion of the
two-orbital model. In Fig.~\ref{fig:2_orb_bands_paramagnetic_11}, the energy
dispersion of the paramagnetic strip is plotted along with
the energies of the extended system projected onto this BZ. We find
energy gaps close to the borders of the BZ. However, there are no edge
bands for the paramagnetic (11) strip. The same holds for the
SDW phase (not shown).

\begin{figure}[t]\centering
\includegraphics[width=0.75\linewidth]{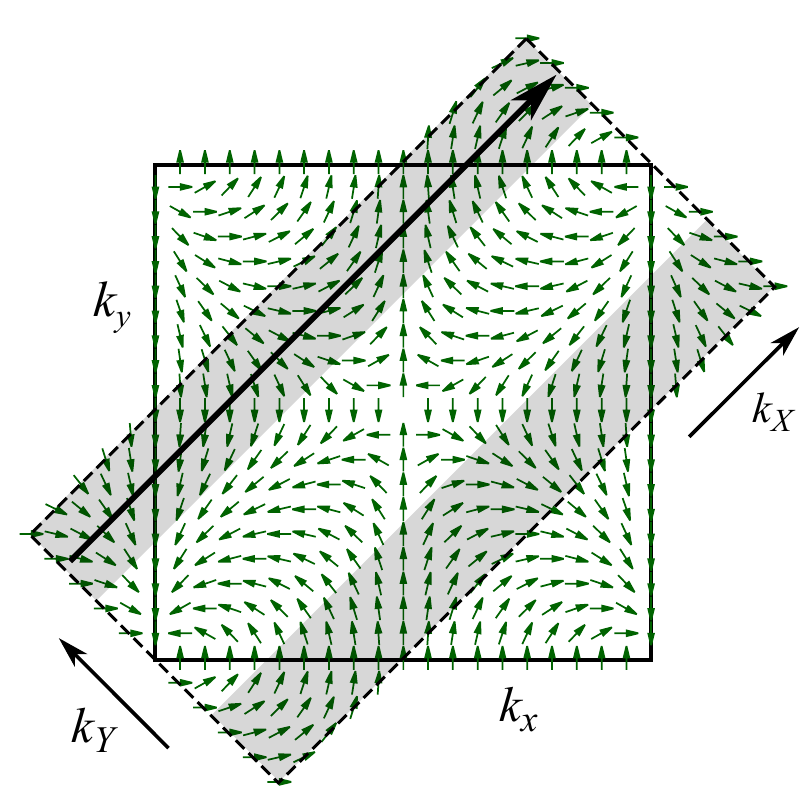}
\caption[Vector field in the extended BZ]
{(Color online) Vector field $(\cos\phi(\mathbf{k}),\sin\phi(\mathbf{k}))$
(green arrows) and exemplary
path at constant $k_Y$ (black arrow) in the extended BZ. The shaded regions
are the areas for which an effective 1D system with fixed $k_Y$ has a bulk gap.
The gap is not necessarily at the Fermi energy.}
\label{fig:vector_field_11}
\end{figure}

We can understand the absence of surface states from a topological
perspective. The argument is similar to the one for the
(10) strip.\cite{LaT13} Following Ran \textit{et al.},\cite{RWZ09} we
rewrite Eq.\ (\ref{eq:2_orb_hamilton_matrix}) as
$\mathcal{H}^0(\mathbf{k}) = a(\mathbf{k})\, \tau^0 + b(\mathbf{k})\,
  [\sin\phi(\mathbf{k})\, \tau^1
  + \cos\phi(\mathbf{k})\, \tau^3]$.
In Ref.\ \onlinecite{LaT13}, we have considered the winding of
$\phi(\mathbf{k})$ at constant $k_y$. For the (11) strip, we have
to analyze paths through the BZ at constant $k_Y$, i.e., diagonal
lines through the BZ associated with the unrotated unit cell, as
illustrated in Fig.~\ref{fig:vector_field_11}.
We see immediately that the winding number for $\phi(\mathbf{k})$
vanishes for all relevant paths. Hence, a continuous deformation of
the Hamiltonian similar to the (10) case, establishing particle-hole and
time-reversal symmetry without
closing the gap, leads to a Hamiltonian in Altland-Zirnbauer class
BDI,\cite{Zir96,SRF08} but with a trivial
topological invariant of $n=0$. Thus, there are no zero-energy end states
in the corresponding finite chain and we do not obtain edge states after
reversing the deformation.

\begin{figure}[t]\centering
\includegraphics[width=0.95\linewidth]{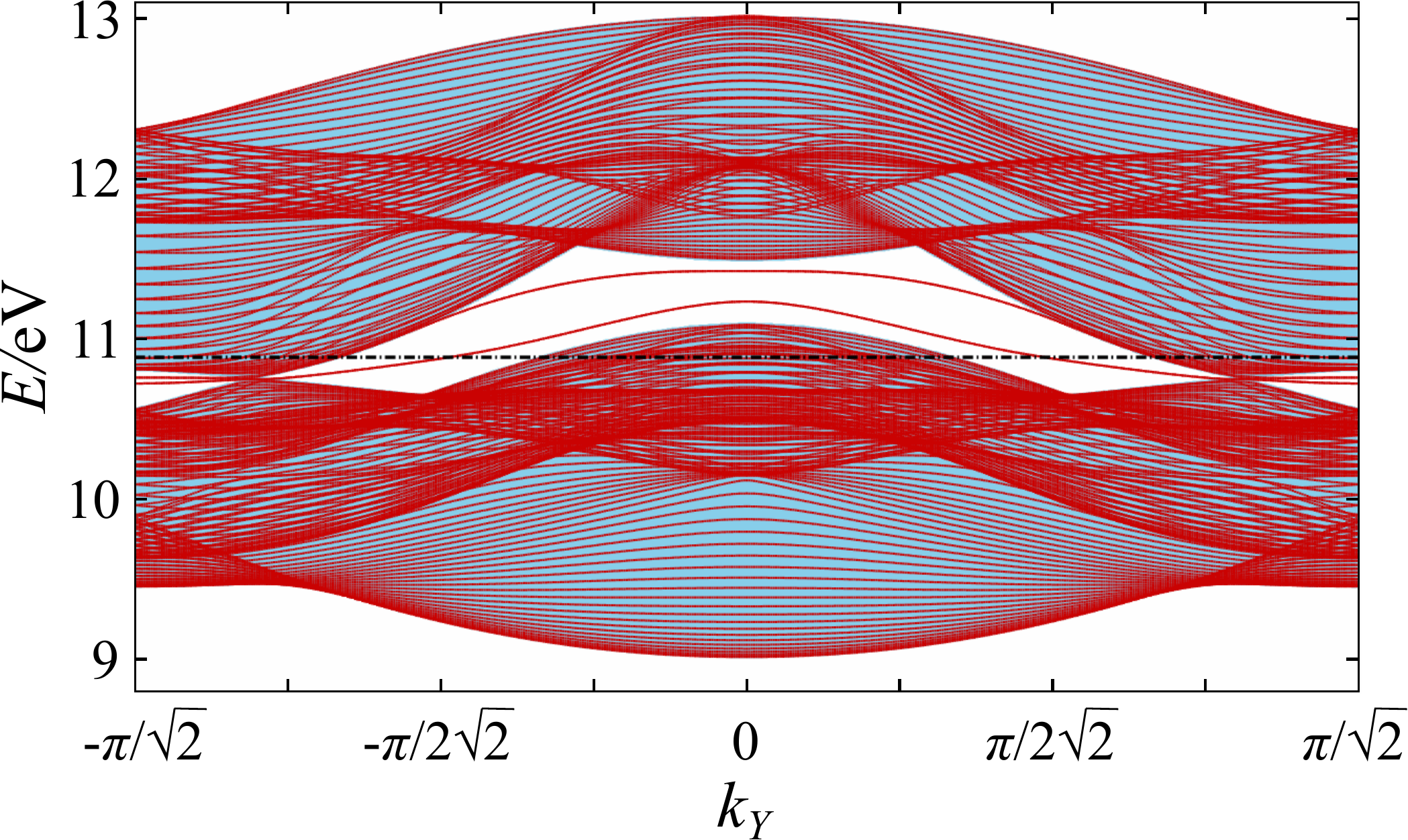}
\caption[Energy bands for the paramagnetic (11) strip in the five-orbital model]
{(Color online) Energy bands in the paramagnetic phase for the
five-orbital model. Bands of a (11) strip of width $W=20$ (red) are
compared to the bulk bands projected onto the
1D BZ for the strip (blue). The black dash-dotted line denotes the Fermi
energy at filling factor $0.6$.}
\label{fig:5_orb_bands_paramagnetic_11}
\end{figure}

For the five-orbital model,
Fig.~\ref{fig:5_orb_bands_paramagnetic_11} shows the band structure of the
system along with the projected bulk spectrum.
Contrary to the results in the two-orbital model, there are
two bundles of surface bands. They are connected to the bulk bands at the
projected Dirac points of the \emph{paramagnetic} system, which do not
exist in the two-orbital model. Like for the (10) edge, each bundle
consists of two pairs of degenerate states with exponentially small
splitting between them for large $W$.

The existence of edge states for the (11) strip can again be understood
based on a topological argument:\cite{LaT13} We obtain effective 1D
Hamiltonians by considering paths through the BZ at constant $k_Y$. Edge
states could in principle exist whenever there is a gap in the bulk spectrum
for this value of $k_Y$. This is the case for all $k_Y$ except
where the path contains Dirac points, at $k_Y \approx \pm 1.8$,
see Fig.\ \ref{fig:5_orb_bands_paramagnetic_11}. There are two classes
of gapped 1D Hamiltonians: the ones for $k_Y$ in the interval spanned
by the projected Dirac points and the ones outside of this interval.
We consider one representative for each class,
corresponding to the paths $\mathcal{C}_1$ at $k_Y=0$ and $\mathcal{C}_2$ at
$k_Y=\pi/\sqrt{2}$, respectively.
All of the following deformations are continuous and do not close the
energy gap. We first decouple the $3d_{3Z^2-R^2}$ orbital
from the others in order to get effective four-orbital Hamiltonians.
We then tune all on-site energies and all hopping amplitudes beyond
next-nearest neighbors to zero. The components of the $4\times 4$ matrices
now consist of linear combinations of $\cos k$, $\cos 2k$, $\sin k$, $\sin 2k$,
and constant terms.

For the Hamiltonian for the path $\mathcal{C}_1$, we continue by
tuning the $\cos 2k$, $\sin 2k$ and the constant terms to zero. After
tuning all remaining coefficients to $1/4$, the resulting matrix is
unitarily equivalent to
\begin{equation}
\hat{\mathcal{H}}_1(k) =
\begin{pmatrix}
0 & e^{-ik} & 0 & 0 \\
e^{ik} & 0 & 0 & 0 \\
0 & 0 & 0 & e^{-ik} \\
0 & 0 & e^{ik} & 0
\end{pmatrix},
\label{eq:5_orb_Hdeformed}
\end{equation}
which consists of two topologically nontrivial two-orbital systems in
class BDI\cite{Zir96,SRF08} with winding numbers $n=1$. This deformed
model has four zero-energy edge bands, two at each edge. These numbers
are doubled if we include the spin. Upon reversing the
deformation, the symmetries defining the class BDI are lost so that the edge
states are no longer required to have zero energy. The edge bands thus become
dispersive. The degeneracy between the two sectors in Eq.\
(\ref{eq:5_orb_Hdeformed}) is also broken and we therefore
end up with two bundles of edge states.

For the path $\mathcal{C}_2$, we tune all nonzero hopping
parameters to the same value denoted by $t$. This is followed by smoothly
tuning the vanishing matrix elements between
$3d_{X^2-Y^2}$ and $3d_{XY}$ to $-2t\cos k$. Next,
the $\cos 2k$, $\sin 2k$, and constant terms are tuned to zero. After
fixing $t$ to $1/2$ and a unitary transformation we obtain the
block Hamiltonian
\begin{equation}
\hat{\mathcal{H}}_2(k) =
\begin{pmatrix}
0 & e^{-ik} & 0 & 0 \\
e^{ik} & 0 & 0 & 0 \\
0 & 0 & 0 & -e^{-ik} \\
0 & 0 & -e^{ik} & 0
\end{pmatrix},
\end{equation}
which comprises two topologically nontrivial two-orbital systems with winding
numbers $n=1$. This deformed system has the same number of zero-energy edge
bands as $\hat{\mathcal{H}}_1(k)$ and the original system thus has two bundles
of edge states also in this $k_Y$ range.

In the antiferromagnetic phase, the 1D surface BZ is halfed due to the
ordering vector $\mathbf{Q}'=(\pi/\sqrt{2},-\pi/\sqrt{2})$ in the rotated
$(k_X,k_Y)$ coordinate system. Hence, the number of bands is doubled and
one would in principle find four bundles of surface bands. However, two of
them become resonant with the bulk states. The degeneracy of the remaining two
bundles at the boundaries of the new BZ is lifted by the SDW. Moreover, we
find that the original four-fold degeneracy for $W\to\infty$ is
still intact, since the magnetization at the (11) edge is staggered. All of
this is similar to the (01) case discussed above.

\subsubsection*{Superconducting phase}

\begin{figure}[t]\centering
\subfloat{\includegraphics[width=0.5\linewidth]
{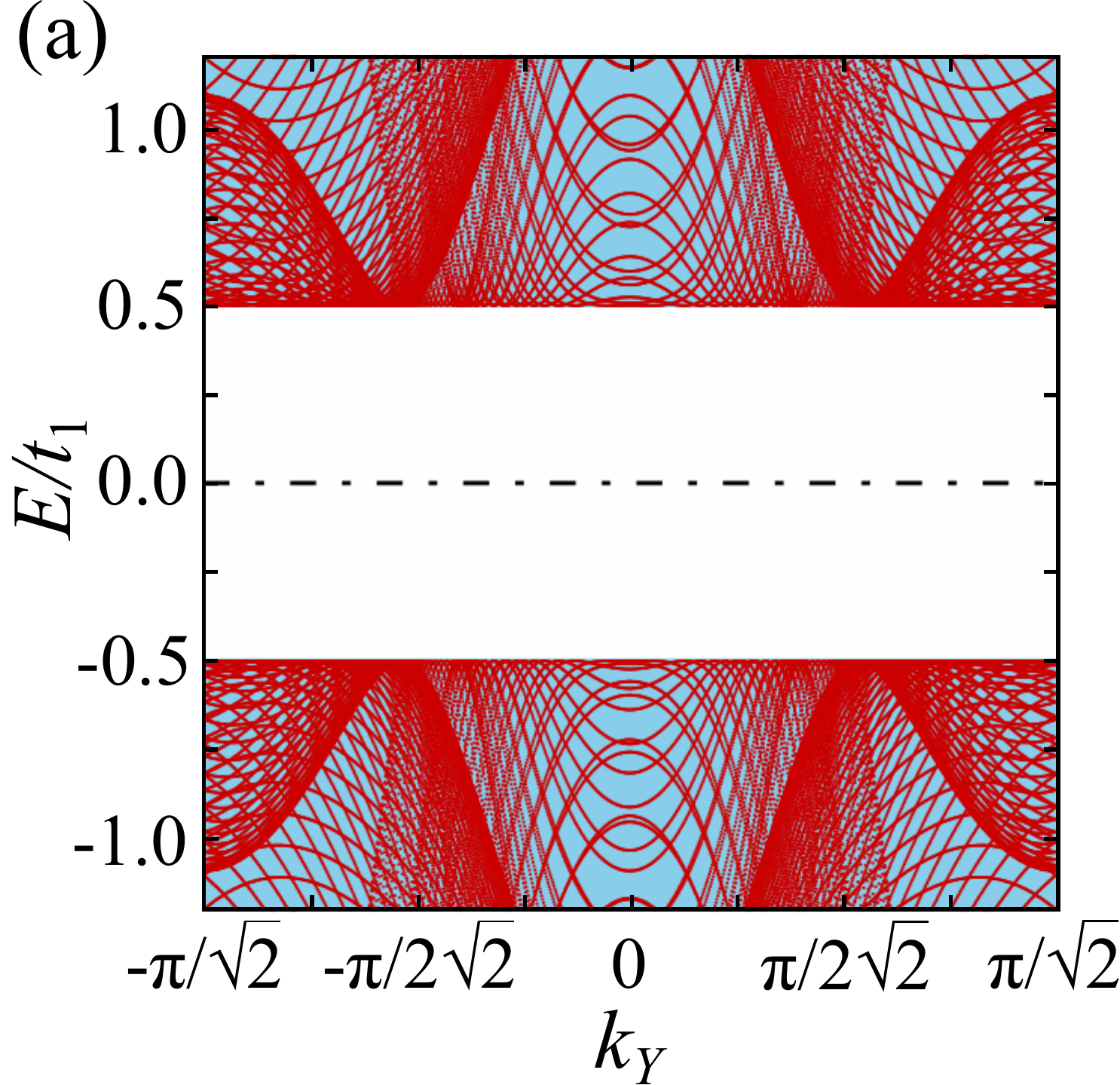}}
\hfill
\subfloat{\includegraphics[width=0.5\linewidth]
{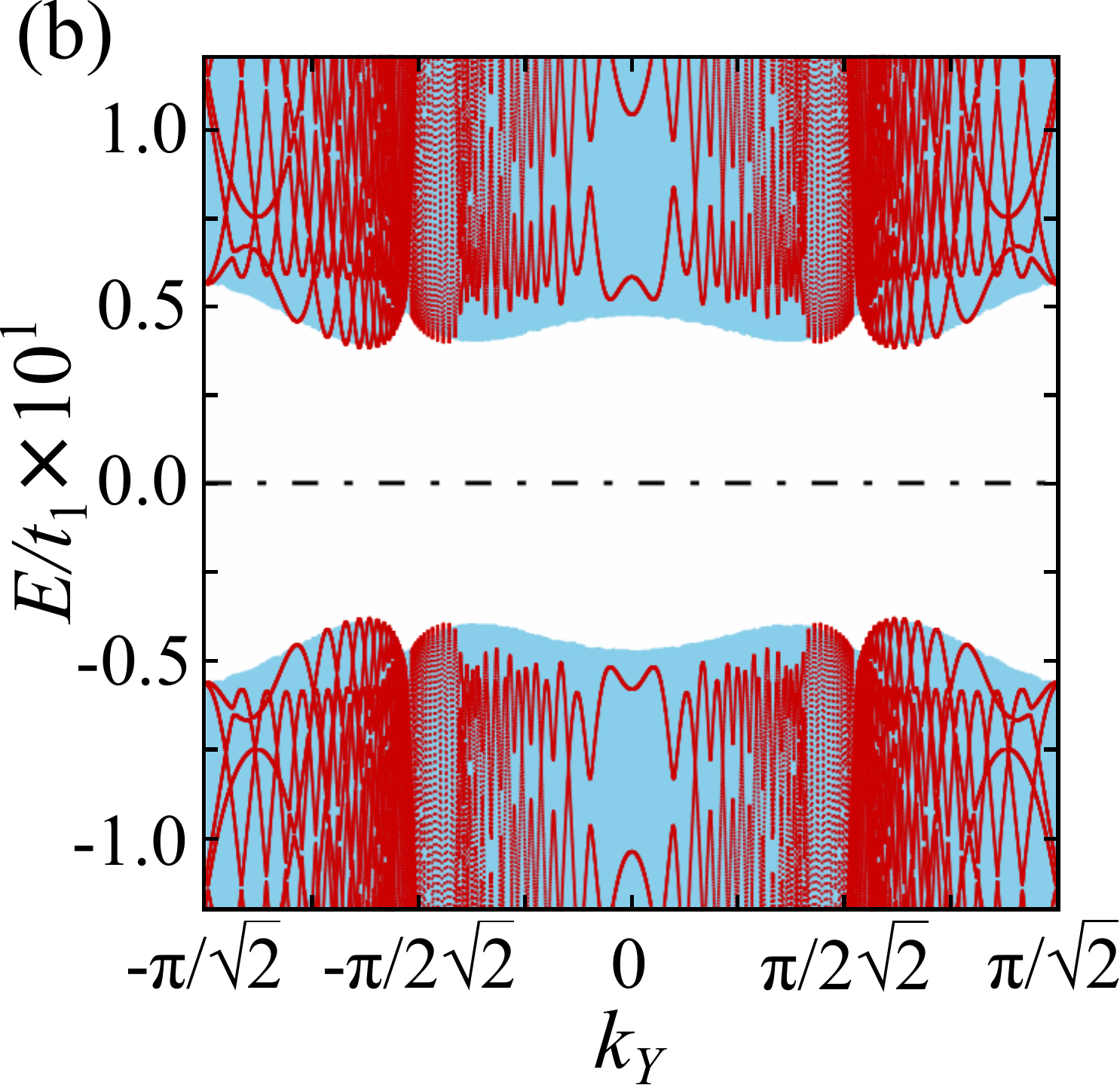}}\\
\vspace{-1em}
\subfloat{\includegraphics[width=0.5\linewidth]
{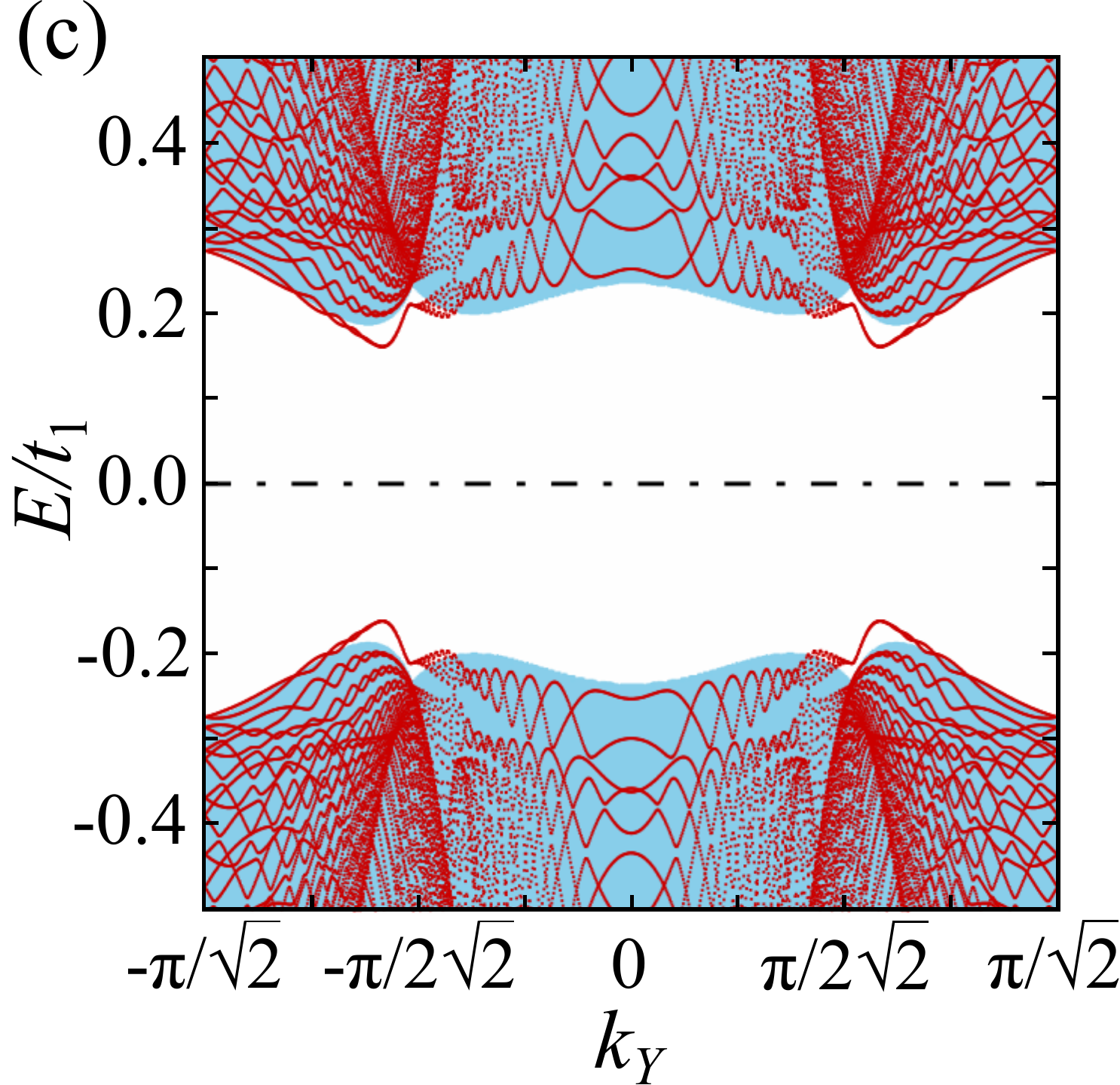}}
\hfill
\subfloat{\includegraphics[width=0.5\linewidth]
{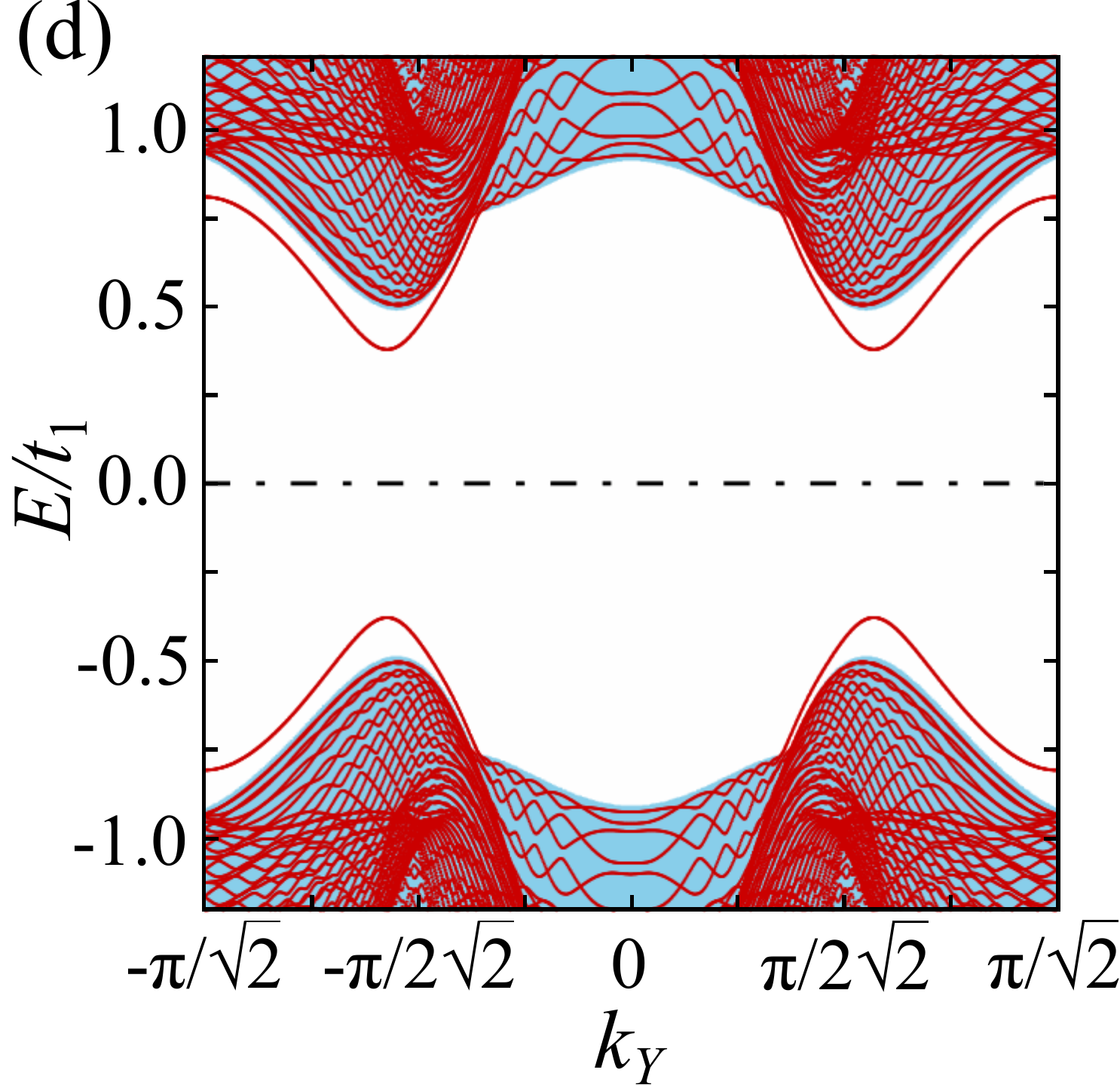}}
\caption[Quasi-particle spectra for the superconducting
(11) strip in the two-orbital model]
{(Color online) Quasi-particle spectra in the superconducting phase for
the two-orbital model. Bands of a (11) strip of width $W=40$ (red) are
compared to the bulk bands projected onto the 1D BZ for the strip (blue).
(a)~$\Delta=0.5$ ($s_{++}$ pairing),
(b)~$\Delta=0.1$ ($s_{\pm}$ pairing),
(c)~$\Delta=0.5$ ($s_{\pm}$ pairing),
(d)~$\Delta=2.0$ ($s_{\pm}$ pairing).
Only the low-energy part of the spectra is shown. Note that there
were no surface states in the normal phase, see Fig.\
\ref{fig:2_orb_bands_paramagnetic_11}.}
\label{fig:2_orb_bands_sc_11}
\end{figure}

For the superconducting phase, we start with the two-orbital model.
In the superconducting phase with $s_{++}$-wave gap,
no surface states appear, see Fig.~\ref{fig:2_orb_bands_sc_11}(a).
This is expected since, on the one hand, this model does not have edge
states at the (11) edge in the normal phase and, on the other, the condition
for the existence of Andreev bound states is not satisfied.

\begin{figure}[t]\centering
\includegraphics[width=1.0\linewidth]{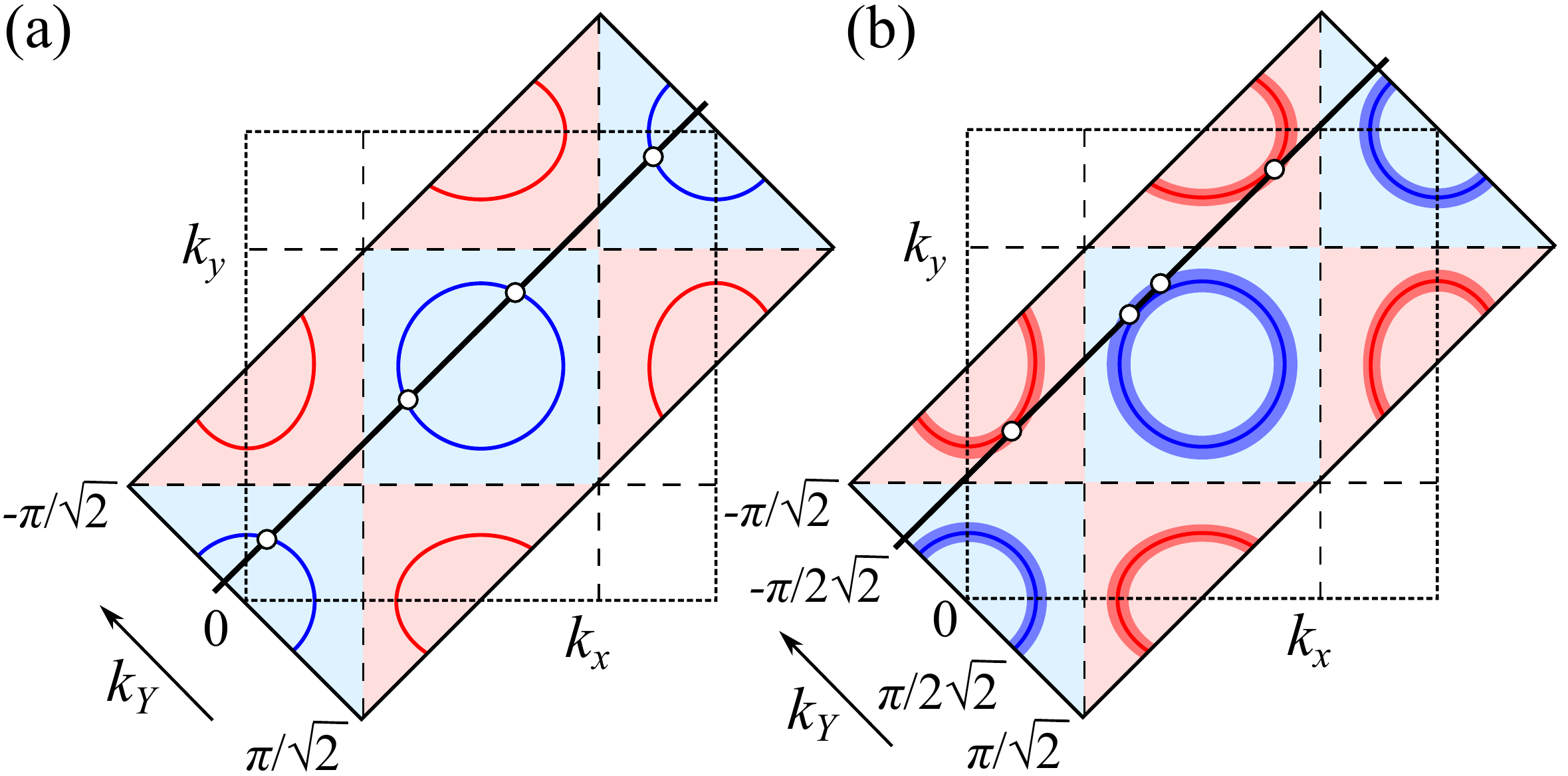}
\caption[Scattering processes for the (11) strip]
{(Color online) Scattering processes in the two-orbital model for the (11)
strip. The extended
BZ is illustrated along with the Fermi surfaces. The bold lines are exemplary
lines with constant $k_Y$. Relevant states are drawn as white circles. (a) Small
$\Delta$: Sign-changing processes can never occur. (b) Larger
$\Delta$:
Relevant states are also found in the vicinity of the Fermi surfaces and
sign-changing scattering processes become possible. Andreev bound states can
emerge.}
\label{fig:sc_scattering_11}
\end{figure}

In the case of $s_\pm$-wave pairing, there are no surface states for
a small gap $\Delta$ as shown in Fig.~\ref{fig:2_orb_bands_sc_11}(b).
This can again be understood from evaluating the sign-changing condition for
Andreev bound states. We consider paths through the extended BZ
at constant $k_Y$, see Fig.~\ref{fig:sc_scattering_11}(a). All
such lines either cross Fermi pockets with the same sign or do not cross a
Fermi surface at all. Hence, sign-changing scattering processes are not possible
if the gap is small and one does not find Andreev bound states. For larger
$\Delta$, also states away from the Fermi surface become relevant, as
indicated in Fig.~\ref{fig:sc_scattering_11}(b). Thus, scattering
processes with a sign change of the gap function can occur,
leading to the emergence of Andreev bound states.
In this case, the Andreev states are
of course not connected to topological states.

\begin{figure}[t]\centering
\subfloat{\includegraphics[width=0.5\linewidth]
{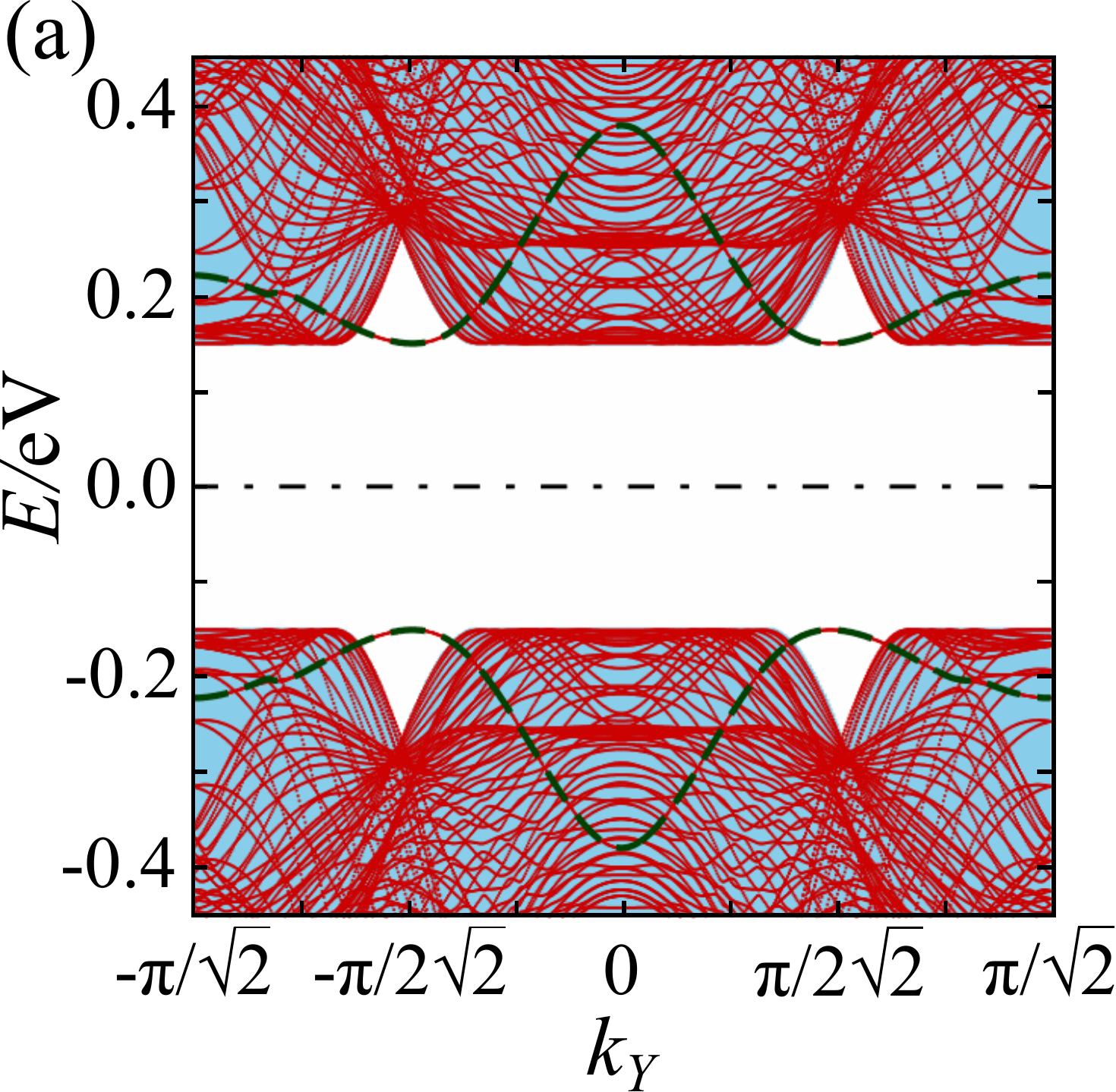}}
\hfill
\subfloat{\includegraphics[width=0.5\linewidth]
{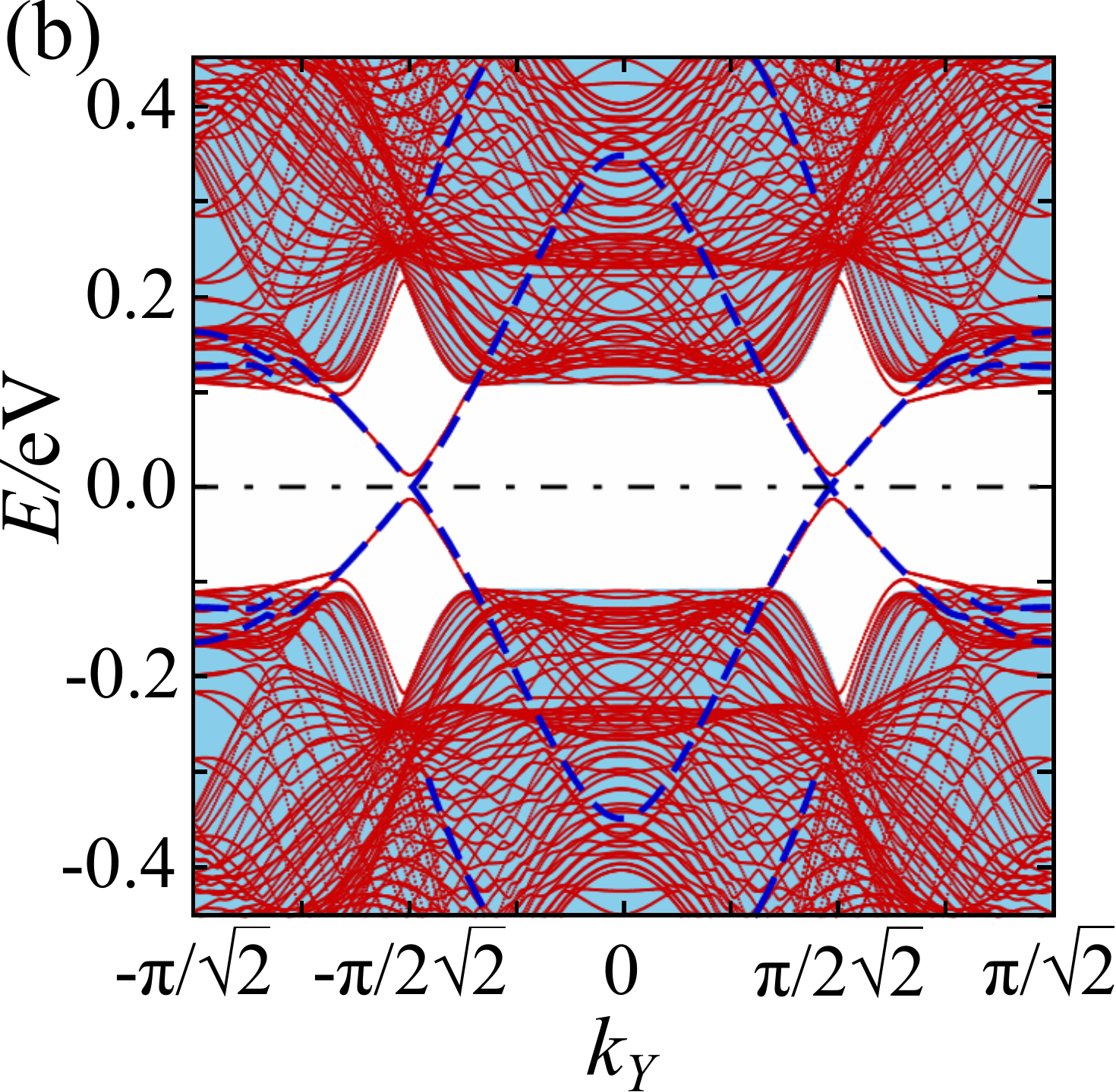}}\\
\caption[Quasi-particle spectra for the superconducting
(11) strip in the five-orbital model]
{(Color online) Quasi-particle spectra in the superconducting phase for the
five-orbital model. Bands of a (11) strip of width $W=20$ (red) are
compared to the bulk bands projected onto the 1D BZ for the strip (blue). Only
the low-energy part of the spectra is shown.
(a)~$\Delta=0.15$ ($s_{++}$ pairing). For comparison, some of the
topological surface bands modified according to $\xi\to\pm\sqrt{\xi^2
+ |\Delta|^2}$ are also plotted (dashed green lines).
(b)~$\Delta=0.15$ ($s_{\pm}$ pairing).
The dashed dark blue lines denote topological surface bands of the
normal state.}
\label{fig:5_orb_bands_sc_11}
\end{figure}

Finally, we turn to the superconducting (11) strip in the five-orbital
model. In Fig.\ \ref{fig:5_orb_bands_paramagnetic_11},
we found two bundles of surface states in the normal phase. In the
superconducting state, the upper bundle vanishes
completely into the bulk continuum, whereas the lower bundle remains partly
in a bulk gap. This is similar to the (10) system in the five-orbital model.
However, the present case is particularly interesting since the
normal-state edge bands cross the Fermi energy.

For $s_{++}$-wave superconductivity, bulk and edge states are again
gapped in the same way, see Fig.~\ref{fig:5_orb_bands_sc_11}(a).
For $s_\pm$-wave pairing, we observe that a larger part of the
lower bundle of topological surface bands remains inside the gap,
hardly affected by the superconducting pairing, see
Fig.~\ref{fig:5_orb_bands_sc_11}(b). However, a very small gap opens,
which is much smaller than the bulk gap. This can again be attributed to
the weakening of the nonlocal $s_\pm$-wave pairing interaction at the edge,
discussed in Sec.\ \ref{sub:1001}.

Furthermore, we find additional surface bands near
$k_Y=\pm\pi/2\sqrt{2}$ very
close to the bulk continuum. These can be understood as Andreev bound states as
discussed for the two-orbital model. We note that the behavior of the
surface bands for larger $\Delta$, is similar to the (10)
strip. The surface band gap grows and the resemblance of the topological bands
to the normal phase gets weaker (not shown). In addition, the Andreev
bands separate more strongly from the bulk continuum.

\section{Conclusions}
\label{sec:con}

We have studied various strip geometries of iron pnictides with
small-index edges in the paramagnetic, antiferromagnetic, and superconducting
phases with regard to the possible existence of surface states. For this,
we have used both a simple two-orbital model\cite{RWZ09} and a more
realistic five-orbital model.\cite{KOA08}

For the paramagnetic phase, we have found that the number of surface bands
depends both on the strip geometry and the specific model considered.
The (10) strip shows edge states in both models.\cite{LaT13} The
two-orbital model predicts one spin-degenerate band of edge states at each edge
(in the limit of large width), resulting from nontrivial winding in the
$3d_{XZ}$, $3d_{YZ}$ orbital space. The five-orbital model has additional
nontrivial winding with regard to the $3d_{X^2-Y^2}$ and $3d_{XY}$ orbitals,
which doubles the number of edge bands.\cite{LaT13} The results for the (11)
strip show that also the winding in the sector of $3d_{XZ}$ and $3d_{YZ}$ is
different between the two models: the two-orbital model is topologically
trivial and thus has no edge states, whereas the five-orbital model has two
spin-degenerate edge bands at each (11) edge. This indicates that the
two-orbital model is too simple to account for the full topological structure
of the pnictide bands.

The presence or absence of surface states can be explained by
considering a continuous deformation of effective 1D Hamiltonians into
Hamiltonians in symmetry class BDI.\cite{Zir96,SRF08,LaT13}
However, these states are no longer topologically protected
and thus move away from the Fermi energy when the deformation is reversed.
It is worth pointing out that this type of argument is rather robust since
it only relies on the existence of a continuous deformation that
does not close a gap. Therefore, the qualitative results, in particular the
existence of surface states, would not change if we included (i) changes in the
model parameters close to the surface, describing possible reconstruction and
relaxation, (ii) order pa\-ra\-me\-ters $m_{ab}$ and $\Delta$ calculated
self-consistently for the strip geometry, or (iii) weak coupling in the third
dimension.

In the antiferromagnetic phase, the degeneracy of the surface bands
in the limit of large width is strongly lifted if the presence of the
SDW leads to a net spin polarization of the edges, which for
$(\pi,0)$ order is the case for $(10)$ edges but not for $(01)$ or $(11)$
edges. Nevertheless, the remaining two-fold degeneracy is protected
by a combination of spin rotation and spatial reflection.

In the superconducting phase with $s_\pm$-wave gap structure, the
topological surface states are less strongly affected by the superconducting
pairing than the bulk states. Only a small gap opens in the surface bands so
that they are almost identical to the normal state and partially
remain inside the gap. In addition, Andreev bound states appear for certain
edges, which can be understood from the changing gap sign for Andreev
reflection.\cite{KaT00,OnT09,NaH09,NHM10,HuL10} For small gaps, the
Andreev bound states coexist with the topological states in different ranges of
the momentum component parallel to the edge. For larger---and for pnictides
unphysical---gap values, the Andreev bound states and topological surface bands
merge and lose their individual character.

The edges studied here correspond to (100), (110), and (010) surfaces in
the real three-dimensional system. However, these surfaces are
challenging to prepare since the natural cleavage plane is (001). More
promising is the examination of single-unit-cell steps on pnictide (001)
surfaces, which are indeed occasionally seen in
scanning-tunneling-microscopy experiments.\cite{Hess}
Since the coupling between layers
in 1111 pnictides is weak, it would only weakly perturb the bound states at
the edge of the incomplete layer. Hence, it should be possible to detect
bound states at step edges with scanning tunneling spectroscopy.
For the detection of bound states in the superconducting phase, it might be
possible to perform tunneling experiments on
normal-superconducting interfaces at the edges of a (001) pnictide sample.

On a more general level, our results emphasize that topological signatures,
such as surface states, can occur in materials that are not
topological in the sense of the topological classification of gapped
systems.\cite{SRF08} Iron
pnictides and graphene are examples of gapless materials that have
topological properties. In the pnictides, the topologically nontrivial
properties come from the multi-orbital character of the band structure close
to the Fermi energy. It is promising to search for other
materials with topological features related to their multi-orbital
structure.

\section*{Acknowledgments}

We thank P. M. R. Brydon and C. Hess for helpful discussions. Support by the
Deutsche Forschungsgemeinschaft through Research Training Group GRK 1621 is
acknowledged.

\end{document}